\documentclass[aps, superscriptaddress, showpacs,pra, twocolumn]{revtex4-1}
\usepackage{ORI_Group_style}
\usepackage[english]{babel}

\usepackage{mathtools}
\usepackage{enumitem}
\usepackage{amsmath,amssymb,amsthm,bm}
\usepackage{xcolor}
\usepackage{array}
\usepackage{mathdots}
\usepackage{hyperref}
\usepackage{upgreek}

\newcommand{\Psiopvec}{\bm{\hat{\Psi}}}

\newcommand{\Phiopvec}{\bm{\hat{\Phi}}}
\newcommand{\Rop}{\hat{R}}
\newcommand{\Ropvec}{\bm{\Rop}}

\newcommand{\Rhopvec}{\bm{\hat{\rho}}}
\newcommand{\Xop}{\hat{X}}

\newcommand{\Mbos}{M_C}
\newcommand{\Mquad}{M}

\newcommand{\Nquad}{N}
\newcommand{\Tbos}{T_C}
\newcommand{\Tquad}{T}

\newcommand{\Kquad}{K}

\newcommand{\KNquad}{K_N}

\newcommand{\gvec}{\bm{g}}
\newcommand{\evec}{\bm{e}}
\newcommand{\fvec}{\bm{f}}
\newcommand{\hvec}{\bm{h}}
\newcommand{\uvec}{\bm{t}}
\newcommand{\vvec}{\bm{s}}
\newcommand{\zvec}{\bm{z}}
\newcommand{\wvec}{\bm{w}}
\newcommand{\xvec}{\bm{x}}
\newcommand{\yvec}{\bm{y}}

\newcommand{\lr}{\mu}               
\newcommand{\li}{\nu}               
\newcommand{\PhiA}{\phi}            
\newcommand{\ThA}{\theta}           
\newcommand{\PhThA}{\gamma}         
\newcommand{\OmegaA}{\alpha}        
\newcommand{\egspace}{\mathcal{H}}  
\newcommand{\lzz}{l_0}              
\newcommand{\nzz}{n_0}              
\newcommand{\lii}{l_i}              
\newcommand{\nii}{n_i}              
\newcommand{\freq}{\eta}            
\newcommand{\coup}{\xi}             
\newcommand{\asign}{\sigma}         

\newcolumntype{L}[1]{>{\raggedright\let\newline\\\arraybackslash\hspace{0pt}}m{#1}}
\newcolumntype{C}[1]{>{\centering\let\newline\\\arraybackslash\hspace{0pt}}m{#1}}
\newcolumntype{R}[1]{>{\raggedleft\let\newline\\\arraybackslash\hspace{0pt}}m{#1}}

\begin{document}

\title{Quadratic Quantum Hamiltonians: \texorpdfstring{\\}{} General Canonical Transformation to a Normal Form}
\author{Katja Kustura}
\affiliation{Institute for Quantum Optics and Quantum Information of the
Austrian Academy of Sciences, A-6020 Innsbruck, Austria.}
\affiliation{Institute for Theoretical Physics, University of Innsbruck, A-6020 Innsbruck, Austria.}
\author{Cosimo C. Rusconi}
\affiliation{Institute for Quantum Optics and Quantum Information of the
Austrian Academy of Sciences, A-6020 Innsbruck, Austria.}
\affiliation{Institute for Theoretical Physics, University of Innsbruck, A-6020 Innsbruck, Austria.}
\author{Oriol Romero-Isart}
\affiliation{Institute for Quantum Optics and Quantum Information of the
Austrian Academy of Sciences, A-6020 Innsbruck, Austria.}
\affiliation{Institute for Theoretical Physics, University of Innsbruck, A-6020 Innsbruck, Austria.}

\begin{abstract}
A system of linearly coupled quantum harmonic oscillators can be diagonalized when the system is dynamically stable using a Bogoliubov canonical transformation. However, this is just a particular case of more general canonical transformations that can be performed even when the system is dynamically unstable. Specific canonical transformations can transform a quadratic Hamiltonian into a normal form, which greatly helps to elucidate the underlying physics of the system. Here, we provide a self-contained review of the normal form of a quadratic Hamiltonian as well as step-by-step instructions to construct the corresponding canonical transformation for the most general case. Among other examples, we show how the standard two-mode Hamiltonian with a quadratic position coupling presents, in the stability diagram, all the possible normal forms corresponding to different types of dynamical instabilities. 
\end{abstract}

\maketitle

\section{Introduction}

Quadratic quantum Hamiltonians ubiquitously appear whenever one describes  coherent quantum dynamics of a system near an  equilibrium point. The quantum fluctuations of every degree of freedom near  equilibrium  can be    described by a quantum harmonic oscillator through a process known as linearization, see \eg~\citep{Rusconi2016}. In this context, the possibility to perform a canonical transformation that diagonalizes the quadratic Hamiltonian, namely a Bogoliubov transformation~\citep{Bogoljubov1958,Valatin1958}, is equivalent to the statement that the system is dynamically stable, see \eg~\citep{Rusconi2017}. Interestingly, whenever the system is dynamically unstable, it is still possible to perform a canonical transformation which, albeit not diagonalizing the Hamiltonian, brings it to a simple form, the so-called normal form~\citep{Williamson1936,Arnold}. 
 
The normal form is defined such that the matrix giving the linear equations of motion of the canonical variables of the system is in a real Jordan normal form, which is a very sparse matrix. The real Jordan normal form of a matrix depends on its spectral properties, in particular, whether it is diagonalizable or not, and on the type of its eigenvalues: real, complex, zero, or imaginary. The Bogoliubov canonical transformation~\citep{Bogoljubov1958,Valatin1958}, for which the normal form Hamiltonian is diagonal, can only be constructed whenever the equation-of-motion matrix is diagonalizable and it has only imaginary eigenvalues. For any of the other many possibilities, a different normal form exists, with a specific canonical transformation, that is associated to a specific type of dynamical instability. For instance, the normal form unveils, among others, which modes of the system  are free particles, which are squeezed, which are interacting via  a beam-splitter type of interaction or via a two-mode squeezing interaction, etc. Thus the normal form of a quadratic Hamiltonian is an enlightening tool to understand the type of quantum many-mode dynamics in  unstable regions as well as to identify the normal dynamical modes of the system.
 
Motivated by this possibility, in this article we revisit previous literature, in particular the results of Laub and Meyer~\citep{Laub1974}, to provide a self-contained material on the normal form of quadratic quantum Hamiltonians as well as detailed explanations on how to construct the canonical transformation for any type of dynamical instability.
In particular, in~\secref{sec:QH} we review the key properties of quadratic quantum Hamtilonians. In~\secref{sec:SpecProp}, we discuss the spectral properties of the equation-of-motion matrix, which  are used in~\secref{sec:NF_Def} to define the normal form of a quadratic quantum Hamiltonian. Step-by-step instructions to construct the generic canonical transformation to transform a quadratic Hamiltonian into its normal form are presented in~\secref{sec:NF_Transf}. Some required technical details are given in App.~\ref{App:gsON}, and the simplified instructions to perform the  Bogoliubov transformation for dynamically stable regimes are reviewed in App.~\ref{App:Bog}. In~\secref{sec:Examples}, we discuss, as an example,  the stability diagram of two harmonic quantum oscillators coupled via their canonical position that presents all types of normal forms associated to different (in)stability regions. We also include the detailed example on how to construct the real canonical transformation to obtain the normal form of a quadratic Hamiltonian describing the interaction  of four coupled quantum harmonic oscillators.  Finally, we draw our conclusions in~\secref{sec:Conclusions}.

\section{Quadratic Hamiltonian}\label{sec:QH}

Let us consider a set of $N$ quantum harmonic oscillators described by the Hermitian operators $\xop_i = \xop_i^\dagger$ and $ \pop_i = \pop_i^\dagger$ for $i=1, \ldots, N$, which are the  dimensionless  position and momentum operators of each oscillator. These operators, which we call quadratures hereafter, satisfy the canonical commutation relations $[\hat x_i, \hat p_j] = \im \delta_{ij}$ and $[\hat x_i, \hat x_j] = [\hat p_i, \hat p_j] = 0$ for $\forall i,j=1, \ldots, N$, where we choose $\hbar=1$ for convenience. A quadratic quantum Hamiltonian is then defined as
\begin{equation}\label{QH}
\hat{H}=\frac{1}{2}\mathbf{\Rop}^T \Mquad \mathbf{\Rop},
\end{equation}
where $ \mathbf{\Ropvec} = (\hat{x}_1,\ldots, \hat{x}_N,\hat{p}_1,\ldots, \hat{p}_N)^T$. The Hamiltonian \eqnref{QH} is  specified by the elements of the real symmetric matrix $ \Mquad = \Mquad^T \in \mathbb{R} ^{2N\times 2N}$, which has units of frequency. We remark that the same matrix $M$ can be used to describe the classical limit of this system~\citep{Note1}.

Instead of writing the quadratic Hamiltonian \eqnref{QH} using quadrature operators, one can use  non-Hermitian creation and annihilation operators defined by $\hat{x}_i=(a_i^{\dagger}+a_i)/\sqrt{2}$ and $\hat{p}_i=\im (a_i^{\dagger}-a_i)/\sqrt{2}$. These operators satisfy bosonic commutation relations $ [\hat a_i, \hat a^\dagger_j] = \delta_{ij}$ and $ [\hat a_i, \hat a_j] =  [\hat a^\dagger_i, \hat a^\dagger_j]=0$ for $\forall i,j=1,\ldots,N$. In this bosonic representation, the Hamiltonian \eqnref{QH} can be written in the quadratic form 
\be 
\Hop=\frac{1}{2} \Psiopvec^\dagger \Mbos \Psiopvec,
\ee 
where $ \Psiopvec = (\hat{a}_1,\ldots, \hat{a}_N,\hat{a}^{\dagger}_1,\ldots, \hat{a}^{\dagger}_N)^T $ and $ \Mbos = \Mbos^\dagger \in \mathbb{C} ^{2N\times 2N}$ is a Hermitian matrix. The quadrature and the bosonic representation are related by $\Ropvec=G\Psiopvec$ and $\Mquad=G \Mbos G^\dagger$, where
\begin{equation}\label{unitary}
\begin{aligned}
G= \frac{1}{\sqrt{2}}\begin{pmatrix} \mathbb{1}_N & \mathbb{1}_N \\ -\im\mathbb{1}_N & \im\mathbb{1}_N  \end{pmatrix}\\
\end{aligned}
\end{equation}
is a unitary matrix and $\mathbb{1}_N$  a $N \times N$ identity matrix.

An important property of a quadratic Hamiltonian is that the system of equations of motion is linear. Indeed, the quadrature Heisenberg equations of motion corresponding to \eqnref{QH} can be written as
\begin{equation}\label{eom}
\frac{\mathrm{d}}{\mathrm{d}t}\Ropvec=J \Mquad \Ropvec \equiv  \Kquad \Ropvec,
\end{equation}
where $J \in \mathbb{R}^{2N \times 2N}$ is an anti-symmetric matrix defined as
\begin{equation}\label{J}
J=\begin{pmatrix} 0 & \mathbb{1}_N \\ -\mathbb{1}_N & 0 \end{pmatrix},
\end{equation}
with the property $J^{-1}=J^T = -J$. In this notation, the commutation rules of the quadratures are given by $[\Rop_i,\Rop_j]=\im J_{ij}$ for $\forall i,j = 1, \ldots, 2N$. The matrix defined as $\Kquad =J \Mquad \in \mathbb{R}^{2N \times 2N} $ is the {\em equation-of-motion matrix} containing all the information about the time evolution of the system (note that $\Mquad = - J \Kquad$). Indeed, \eqnref{eom} can be formally integrated to 
\be \label{eq:Rvst}
\Ropvec(t)=\exp(\Kquad t)\Ropvec(0).
\ee
At this point, it is clear that the dynamical stability of the system, characterized by the fact that none of the mean values $|\avg{ \Rop_i (t)}|$, $|\avg{ \Rop_i (t) \Rop_j(t)}|$ for $\forall i,j =1,\ldots, 2N$ grow indefinitely as a function of time, depends on the spectral properties of the equation-of-motion matrix $\Kquad$. These properties are dictated by the defining condition of $\Kquad$:
\begin{equation}\label{Kproperty}
J \Kquad + \Kquad^T J = 0,
\end{equation}
which guarantees that commutation relations are preserved during the time evolution, namely $[\Rop_i(t),\Rop_j(t)]=\im J_{ij}$ for $\forall i,j = 1, \ldots, 2N$. As further discussed in \secref{sec:SpecProp}, the property \eqnref{Kproperty} allows $\Kquad$ to have real, complex, zero, or imaginary eigenvalues, as well as to be diagonalizable or non-diagonalizable. As we will show later, dynamical stability corresponds to $\Kquad$ being diagonalizable and only having imaginary eigenvalues~\citep{Note2}.

As commonly done in mechanics, one can find a new set of coordinates in which the system can be more conveniently described. That is, one can perform a change of coordinates via a $2N \times 2N$ transformation matrix $\Tquad$, such that $\Ropvec=\Tquad \Rhopvec$, where $ \mathbf{\Rhopvec} = (\hat{X}_1,..., \hat{X}_N,\hat{P}_1,..., \hat{P}_N)^T$ are the new coordinates. In order to guarantee that the new operators are Hermitian $\hat \rho _i^\dagger=\hat \rho_i$ and fulfill canonical commutation rules $[\hat{\rho}_i,\hat{\rho}_j]  =  \im J_{ij}$ for $\forall i,j = 1, \ldots, 2N$, then the so-called {\em real canonical transformation} $T$ has to
\begin{enumerate}
    \item be real $\Tquad \in \mathbb{R}^{2N \times 2N}$ and
    \item satisfy $\Tquad J \Tquad^T = J$, which is known as the symplectic condition.
\end{enumerate}
The Hamiltonian in terms of the new quadratures is then
\begin{equation}\label{QH-NF}
\begin{aligned}
\hat{H}= \frac{1}{2} \Rhopvec^T \Nquad \Rhopvec,
\end{aligned}
\end{equation}
with $\Nquad=\Tquad^T \Mquad \Tquad$. The equations of motion for $\Rhopvec$ have the same structure as \eqnref{eom}, with the equation-of-motion matrix $\KNquad = J \Nquad$, which is obtained from $\Kquad$ via 
\begin{equation}\label{similarity}
\KNquad = \Tquad^{-1} \Kquad \Tquad.
\end{equation}
The transformation of the equation-of-motion matrix $\Kquad$ \eqnref{similarity}, called a similarity transformation, is the one used to diagonalize a diagonalizable matrix by rewriting it in an eigenbasis. Note, however, that even if $\Kquad$ is diagonalizable via a similarity transformation, it does not mean that this can be achieved with a real canonical transformation.
Finally, the canonical transformation in the bosonic representation is given by $\Tbos = G^\dagger \Tquad G$, with $\Psiopvec=\Tbos \Phiopvec$, where $ \Phiopvec = (\hat{b}_1,..., \hat{b}_N,\hat{b}^{\dagger}_1,..., \hat{b}^{\dagger}_N)^T$ are the new bosonic operators with $[ \hat{b}_i, \hat{b}_j^\dagger] = \delta_{ij}$ and $ [\hat b_i, \hat b_j] =  [\hat b^\dagger_i, \hat b^\dagger_j]=0$ for $\forall i,j=1,\ldots,N$. 

The goal is to construct a real canonical transformation that gives an equation-of-motion matrix $\Kquad$ that is as sparse as possible. How to achieve this depends on the spectral properties of $\Kquad$, which we discuss in the following \secref{sec:SpecProp}.

\section{Spectral properties of the equation-of-motion matrix \texorpdfstring{$\Kquad$}{K}} \label{sec:SpecProp}

In this section, the spectral properties of the equation-of-motion matrix $\Kquad$ are analyzed and the different possibilities classified. This information is used to define the normal form of a quadratic Hamiltonian (\secref{sec:NF_Def}) as well as to explain how to construct real canonical transformations that bring a general quadratic quantum Hamiltonian to its normal form (\secref{sec:NF_Transf}).

\subsection{Eigenvalues of \texorpdfstring{$\Kquad$}{K}}
The equation-of-motion matrix $\Kquad \in \mathbb{R}^{2N\times 2N}$ has complex eigenvalues that we denote as $\lambda_i$. A given eigenvalue $\lambda_i$ has an algebraic multiplicity $a_i$ (the number of times $\lambda_i$ is the root of the characteristic polynomial of $\Kquad$) and a geometric multiplicity $m_i$ (the number of linearly independent eigenvectors of $\lambda_i$, see \secref{GEVs}).  We remark that $m_i \leq a_i$, and the matrix $\Kquad$ is diagonalizable if and only if $m_i = a_i$ for all eigenvalues $\lambda_i$~\citep{Horn}.

The defining condition  \eqnref{Kproperty}  restricts the form of $\Kquad$ to
\begin{equation}\label{Kquad}
\Kquad = \begin{pmatrix} A_I & A_R \\ A_L & -A_I^T  \end{pmatrix},
\end{equation}
where $A_I,A_R,A_L \in \mathbb{R}^{N\times N} $ and $ A_R=A_R^T$, $A_L=A_L^T$~\cite{Broadbridge1979}. More importantly, it restricts the complex eigenvalues $\lambda_i$ of $\Kquad$ in the following way \cite{Meiss2007}: 
\begin{itemize}
    \item if  $\lambda_i$ is non-zero and real, $-\lambda_i$ is also an eigenvalue with the same multiplicities $a_i$ and $m_i$. We call $\mathcal{R}_i = \lbrace \lambda_i, -\lambda_i\rbrace$ a real pair and $N_\mathcal{R}$ the total number of real pairs.
    \item if  $\lambda_i$ has a non-zero real and imaginary part, $\bar\lambda_i$, $-\lambda_i$, and $-\bar\lambda_i$ are also eigenvalues with the same multiplicities $a_i$ and $m_i$. We call $\mathcal{C}_i = \lbrace \lambda_i, -\lambda_i,\bar \lambda_i,-\bar\lambda_i\rbrace$ a complex quadruplet  and $N_\mathcal{C}$ the total number of complex quadruplets.
    \item if $\lambda_i=0$, its algebraic multiplicity $a_i$ is even. Throughout this article we will assign the label $i=0$ to the zero eigenvalue, namely $\lambda_i=0$ if and only if $i=0$.
    \item if  $\lambda_i$ is non-zero and purely imaginary, $\bar\lambda_i=-\lambda_i$ is also an eigenvalue with the same multiplicities $a_i$ and $m_i$. We call $\mathcal{I}_i = \lbrace \lambda_i, \bar\lambda_i\rbrace$ an imaginary pair  and $N_\mathcal{I}$ the total number of imaginary pairs.
\end{itemize}
Note that by summing the degeneracies of the zero case, every real pair, every complex quadruplet, and every imaginary pair, one has that
\be 
2 N = a_0 + 2 \sum_{\mathcal{R}_i} a_i + 4 \sum_{\mathcal{C}_i} a_i + 2 \sum_{\mathcal{I}_i} a_i.
\ee

\subsection{Generalized eigenvectors of \texorpdfstring{$\Kquad$}{K}}\label{GEVs}
For every eigenvalue $\lambda_i$ of $\Kquad$ there are $m_i$ eigenvectors, which we denote as $\gvec_{ij}^{(1)} \in \mathbb{C}^{2N}$, for $j=1,\dots, m_i$. The eigenvectors span an $m_i$-dimensional eigenspace, which we denote as $\egspace^{(1)}(\lambda_i)$. In case $m_i < a_i$, one can introduce generalized eigenvectors (GEVs) $\gvec_{ij}^{(k)}$, which are defined by~\citep{Olver}
\begin{equation}
\begin{aligned}
( \Kquad - \lambda_i \mathbb{1})^k \gvec_{ij}^{(k)} = 0,\\ 
( \Kquad - \lambda_i \mathbb{1})^{k-1} \gvec_{ij}^{(k)} \neq 0,
\end{aligned}
\end{equation}
where $\mathbb{1}\equiv \mathbb{1}_{2N}$. The superindex $k$ labels the rank of the GEV and hence a GEV with $k=1$ is an eigenvector. For a given eigenvalue $\lambda_i$ and an eigenvector $\gvec_{ij}^{(1)}$, there are $D_{ij}$ linearly independent GEVs: $\lbrace \gvec_{ij}^{(1)}, \dots, \gvec_{i j}^{(D_{ij}-1)}, \gvec_{i j}^{(D_{ij})} \rbrace$, that form a so called Jordan chain of length  $D_{ij}$. A Jordan chain can be constructed from the generating GEV (gGEV), which is the GEV of the highest rank that we denote for later convenience as $\gvec_{ij} \equiv \gvec_{ij}^{(D_{ij})}$, by~\citep{Olver}
\begin{equation}
\gvec^{(k)}_{ij} = ( \Kquad - \lambda_i \mathbb{1})^{D_{ij}-k} \gvec_{ij}, \quad k=1,\dots,D_{ij}.
\end{equation}
The algebraic multiplicity of an eigenvalue $\lambda_i$ is given by the sum of all its Jordan chain lengths, namely 
\be
a_i = \sum_{j=1}^{m_i} D_{ij}.
\ee
The GEVs in all the Jordan chains of an eigenvalue $\lambda_i$ span an $a_i$-dimensional generalized eigenspace that we denote as $\egspace(\lambda_i)$. Hence, the eigenspace spanned by the eigenvectors with eigenvalue $\lambda_i$, namely $\egspace^{(1)}(\lambda_i)$, is a subspace of $\egspace(\lambda_i)$.
When $K$ is diagonalizable, all the Jordan chains contain only one element ($D_{ij}=1$ $\forall i,j$), and thus $\gvec^{(1)}_{ij}=\gvec_{ij}$.

For the discussion of the normal form in the following sections, it is convenient to make some definitions and introduce some notation:
\begin{itemize}
    \item Real pair $\mathcal{R}_i=\cpare{\lambda_i,-\lambda_i}$: there are $m_i$ gGEVs $\gvec_{ij} \in {\egspace}(\lambda_i)$ and $m_i$ gGEVs $ \tilde \gvec_{ij} \in {\egspace}(-\lambda_i)$.
    \item Complex quadruplet  $\mathcal{C}_i=\cpare{\lambda_i, - \lambda_i, \bar \lambda_i, - \bar \lambda_i}$: there are $m_i$ gGEVs $\gvec_{ij} \in {\egspace}(\lambda_i)$ and $m_i$ gGEVs $ \tilde \gvec_{ij} \in {\egspace}(- \lambda_i)$. Consequently, there are also $m_i$ gGEVs $\bar{{ \gvec}}_{ij} \in {\egspace}(\bar \lambda_i)$ and $m_i$ gGEVs $\bar{\tilde{ \gvec}}_{ij} \in {\egspace}(- \bar \lambda_i)$.
    \item Zero $\cpare{\lambda_0=0}$: there are $m_0$ gGEVs $\gvec_{0j} \in {\egspace}(0)$.
    \item Imaginary pair $\mathcal{I}_i=\cpare{\lambda_i, \bar \lambda_i}$: there are $m_i$ gGEVs $\gvec_{ij} \in {\egspace}(\lambda_i)$ and $m_i$ gGEVs $ \tilde \gvec_{ij} = \bar \gvec_{ij} \in {\egspace}(\bar \lambda_i)$.
\end{itemize}
We remark that by a proper ordering, one has that for any $j=1,\ldots,m_i$ both $\gvec_{ij}$ and $\tilde \gvec_{ij}$ have the same rank $D_{ij}$. 

It is  convenient to define the complex number
\begin{equation}\label{alpha}
\alpha_{\lambda}(\xvec, \tilde \yvec) \equiv \spare{(K-\lambda \mathbb{1})^{D-1} \xvec}^T J \tilde \yvec 
\end{equation}
where $\xvec \in \egspace(\lambda)$ is a gGEV of rank $D$ and $\tilde \yvec \in \egspace(-\lambda)$. In the definition \eqnref{alpha}, $\xvec$ and $\tilde \yvec$ do not have to be of the same rank and for zero eigenvalue $\lambda_0=0$ it should be understood that $\tilde \yvec  \in \egspace(0)$. \eqnref{alpha} can be understood as a special type of a scalar product of the form $\bm{v}^T J \bm{u}$, where $\bm{v} = (K-\lambda \mathbb{1})^{D-1} \xvec$ and $\bm{u} = \tilde \yvec$. Indeed, this form is called the standard symplectic inner product~\citep{Laub1974,Meyer}. It arises naturally in this context because the commutation relations of the quadratures are defined by the matrix \eqnref{J}. It is always possible to transform a set of gGEVs $\gvec_{ij}$ to a new set of gGEVs $\evec_{ij}$ with the same $D_{ij}$ such that they fulfill the following properties~\citep{Laub1974,Meyer}: 
\begin{itemize}
    \item Real pair $\mathcal{R}_i=\cpare{\lambda_i,-\lambda_i}$: one transforms $\gvec_{ij} \rightarrow \evec_{ij}$ and $\tilde \gvec_{ij} \rightarrow \tilde \evec_{ij}$ such that $\alpha_{\lambda_i}(\evec_{ij},\tilde \evec_{ij'})= \delta_{jj'}$.
    \item Complex quadruplet  $\mathcal{C}_i=\cpare{\lambda_i, - \lambda_i, \bar \lambda_i, - \bar \lambda_i}$: one transforms $\gvec_{ij} \rightarrow \evec_{ij}$ and $\tilde \gvec_{ij} \rightarrow \tilde \evec_{ij}$ such that $\alpha_{\lambda_i}(\evec_{ij},\tilde \evec_{ij'})= \delta_{jj'}$.
    \item Zero $\cpare{\lambda_0=0}$: one transforms $\gvec_{0j} \rightarrow \evec_{0j}$ such that 
      \be
    \begin{split}
        \alpha_0(\evec_{0j},\evec_{0j})= 
        \begin{cases}
        \pm 1, & D_{0j}=\mathrm{even},\\
        0, & D_{0j}=\mathrm{odd}.
        \end{cases}
    \end{split}
    \ee
    When $j \neq j'$, then $\alpha_0(\evec_{0j},\evec_{0j'}) =0$ if $D_{0j}$ or $D_{0j'}$ is even, otherwise $\alpha_0(\evec_{0j},\evec_{0j'})$ can be different from zero.
    \item Imaginary pair $\mathcal{I}_i=\cpare{\lambda_i, \bar \lambda_i}$: one transforms $\gvec_{ij} \rightarrow \evec_{ij}$ such that $\alpha_{\lambda_i}(\evec_{ij},\bar \evec_{ij'})= \delta_{jj'} \asign_{ij}$, where if $D_{ij}$ is even (odd), then $\asign_{ij}=\pm 1$ ($\asign_{ij}=\pm \im$).
\end{itemize}
The transformation $\gvec_{ij} \rightarrow \evec_{ij}$ can be understood as an orthonormalization of the gGEVs with respect to the form \eqnref{alpha}. How to perform this transformation, which we call generalized symplectic orthonormalization, is explained in \appref{App:gsON}. Hereafter, we assume that the gGEVs are in the symplectic orthonormalized form $\evec_{ij}$.

At this point, one can make a classification of the gGEVs in the symplectic orthonormalized form $\evec_{ij}$ that will be used to define the normal form in \secref{sec:NF_Def} as well as the general canonical transformation in \secref{sec:NF_Transf}. We define six cases denoted by $\mathfrak{c}=1,\ldots,6$:
\begin{enumerate}
    \item $\evec_{ij}$ belongs to case $\mathfrak{c}=1$ if $\evec_{ij}$ is the gGEV of a real pair $\mathcal{R}_i$. For each $\mathcal{R}_i$, there are $m_i$ gGEVs $\evec_{ij}$ of the type $\mathfrak{c}=1$.
    \item$ \evec_{ij}$ belongs to case $\mathfrak{c}=2$ if $\evec_{ij}$ is the gGEV of a complex quadruplet $\mathcal{C}_i$. For each $\mathcal{C}_i$, there are $m_i$ gGEVs $\evec_{ij}$ of the type $\mathfrak{c}=2$.
    \item $\evec_{0j}$ belongs to case $\mathfrak{c}=3$ if $\evec_{0j}$ is the gGEV of a zero eigenvalue with an even rank $D_{0j}$. There are $\lzz$ gGEVs $\evec_{0j}$ of the type $\mathfrak{c}=3$ and they have $\alpha_0 (\evec_{0j},\evec_{0j})=\pm 1$.
    \item $\evec_{0j}$ belongs to case $\mathfrak{c}=4$ if $\evec_{0j}$ is the gGEV of a zero eigenvalue with an odd rank $D_{0j}$. There are $2\nzz$ gGEVs $\evec_{0j}$ of the type $\mathfrak{c}=4$ and they have $\alpha_0 (\evec_{0j},\evec_{0j})=0$. Note that $m_0 = \lzz+2\nzz$.
    \item $\evec_{ij}$ belongs to case $\mathfrak{c}=5$ if $\evec_{ij}$ is the gGEV of an imaginary pair $\mathcal{I}_i$ with an even rank $D_{ij}$. For each $\mathcal{I}_i$, there are $\lii$ gGEVs $\evec_{ij}$ of the type $\mathfrak{c}=5$ and they have $\alpha_{\lambda_i} (\evec_{ij},\bar \evec_{ij})=\pm 1$.
    \item $\evec_{ij}$ belongs to case $\mathfrak{c}=6$ if $\evec_{ij}$ is the gGEV of an imaginary pair $\mathcal{I}_i$ with an odd rank $D_{ij}$. For each $\mathcal{I}_i$, there are $\nii$ gGEVs $\evec_{ij}$ of the type $\mathfrak{c}=6$ and they have $\alpha_{\lambda_i} (\evec_{ij},\bar \evec_{ij})=\pm \im$. Note that $m_i = \lii+\nii$.
\end{enumerate}
Let us make a remark about the notation used in this article. In indexing the gGEVs, we do not specify the exact case they belong to in order to ease the notation. Whenever we refer to a particular eigenvalue $\lambda_i$ and its corresponding case $\mathfrak{c}$, then it should be clear that symbols such as $\evec_{ij}$ and $D_{ij}$ denote the gGEVs and ranks belonging to this case. 

At this point, we have all the ingredients to define the normal form of a quadratic Hamiltonian given in \secref{sec:NF_Def}. 

\section{Normal Form Definition}\label{sec:NF_Def}

We define the normal form of a quadratic Hamiltonian as the form obtained by a real canonical transformation $T$ such that the equation-of-motion matrix $\Kquad$ is in a real Jordan normal form \cite{Arnold}. 

Using all the ingredients and notation discussed in \secref{sec:SpecProp}, one has that the real Jordan normal form of $\Kquad$ is 
\begin{equation}\label{KN}
\KNquad = \begin{pmatrix} O_I & O_R \\ O_L & -O_I^T  \end{pmatrix},
\end{equation}
where $O_\chi \in \mathbb{R}^{N\times N} $ ($\chi=I,L,R$), $O_R= O_R^T$, and $O_L= O_L^T$. The matrices $O_\chi$ can each be expressed as a direct sum of blocks over different eigenvalue types
\be
O_\chi =  O_\chi^{(\mathfrak{\mathcal{R}})} \oplus O_\chi^{(\mathfrak{\mathcal{C}})} \oplus O_\chi^{(0)} \oplus O_\chi^{(\mathfrak{\mathcal{I}})},
\ee
where
\begin{eqnarray}
O_\chi^{(\mathfrak{\mathcal{R}})} &=& \bigoplus_{\mathcal{R}_i   }  \spare{\Moplus_{j=1}^{m_i} I_{\chi}^{(1)}(\lambda_i,D_{ij})},\\
O_\chi^{(\mathfrak{\mathcal{C}})} &=& \bigoplus_{\mathcal{C}_i }  \spare{\Moplus_{j=1}^{m_i} I_{\chi}^{(2)}(\lambda_i,D_{ij})},\\
O_\chi^{(0)} &=&  \bigoplus_{j=1}^{\lzz} I_{\chi}^{(3)}(0,D_{0j}) \bigoplus_{j=1}^{\nzz} I_{\chi}^{(4)}(0,D_{0j}),\\
O_\chi^{(\mathcal{I})} &=& \bigoplus_{\mathcal{I}_i  }  \spare{ \Moplus_{j=1}^{\lii} I_{\chi}^{(5)}(\lambda_i,D_{ij}) \Moplus_{j=1}^{\nii} I_{\chi}^{(6)}(\lambda_i,D_{ij})}.
\end{eqnarray}
The 18 matrices $I^{(\mathfrak{c})}_\chi (\lambda_i, D_{ij})$ for $\mathfrak{c}=1,\ldots,6$ and $\chi=I,L,R$ are given in \tabref{TAB:NormalFormList}. Note that eight out of the 18 matrices are zero. 
\begin{table*}
	\caption{Blocks for the real Jordan normal form of the equation-of-motion matrix $\Kquad$. We denote $\lambda = \lr + \im \li$ (with  $\lr,\li \in \mathbb{R}$) and $\asign=\alpha_{\lambda}(\evec, \tilde \evec)$, where $\evec$ is the generating generalized eigenvector (gGEV) associated to the block. Note that in $\mathfrak{c}=6$, $\im\sigma \in \mathbb{R}$.}\label{TAB:NormalFormList}
	\begin{tabular}{|C{0.04\textwidth} | C{0.28\textwidth} | C{0.28\textwidth} |C{0.28\textwidth} | C{0.09\textwidth} |}
		\hline
		$\mathfrak{c}$ & $I_I^{(\mathfrak{c})} (\lambda, D)$ & $I_R^{(\mathfrak{c})}(\lambda,D)$ & $I_L^{(\mathfrak{c})}(\lambda,D)$ & dimension \\
		\hline
		1 
		&
		\newline
		$\begin{pmatrix}
		\lambda &  &  &   &  \\
		1 & \lambda &  &  &  \\ 
		& \ddots & \ddots &  &  \\ 
		&  & & 1 & \lambda \\
		\end{pmatrix}$ 
		
		&
		$\mathbb{0}$
		&
		$\mathbb{0}$
		&
		$D$
		\\
		\hline
		2
		&
		\newline
		$\begin{pmatrix}
		\lr & \li &  &  &  &  &  &  \\
		-\li & \lr &  &  &  &  &  &  \\
		1 & 0 & \lr & \li &  &  &  &  \\
		0 & 1 & -\li & \lr &  &  &  &  \\
		&  &  & \ddots & \ddots &   &  &  \\
		&  &  &  & 1 & 0 & \lr & \li \\
		&  &  &  & 0 & 1 & -\li & \lr \\
		\end{pmatrix}$
		
		&
		$\mathbb{0}$
		&
		$\mathbb{0}$
		&
		$2D$
		\\
		\hline
		3
		&
		\newline
		$\asign
		\begin{pmatrix}
		0 &  &  &   \\ 
		1 & 0 &  &  \\ 
		& \ddots & \ddots & \\ 
		&   & 1 & 0\\
		\end{pmatrix} $
		
		&
		$\mathbb{0}$
		&
		$\asign
		\begin{pmatrix}
		0 &  &  &   \\ 
		& 0 &  &  \\ 
		&  & \ddots & \\ 
		&   &  & (-1)^{D/2}\\
		\end{pmatrix} $
		&
		$D/2$
		\\
		\hline
		4
		&
		\newline
		$\begin{pmatrix}
		0 &  &  &   \\ 
		1 & 0 &  &  \\ 
		& \ddots & \ddots & \\ 
		&   & 1 & 0\\
		\end{pmatrix} $
		
		&
		$\mathbb{0}$
		&
		$\mathbb{0}$
		&
		$D$ (odd)
		\\
		\hline
		5
		&
		\newline
		$\mathbb{0}$
		
		&
		\newline
		$\asign
		\begin{pmatrix}
		&  & &  &  & \li\\
		&  &  & & \li & 1\\
		&  &  & & -1 & \\
		&  &\iddots  & \iddots &  & \\
		& \li & -1 &  & & \\
		\li & 1 &  & &  & \\
		\end{pmatrix}$
		
		& 
		$\asign
		\begin{pmatrix}
		&  & &  &  -1 & -\li\\
		&  &  & 1& -\li & \\
		&  &\iddots  & \iddots &  & \\
		& 1 &  & & & \\
		-1 & -\li &  &  & & \\
		-\li & &  & &  & \\
		\end{pmatrix}$
		&
		$D$ (even)
		\\
		\hline
		6
		&
		\newline
		$\begin{pmatrix}
		0 &  &  &   \\ 
		1 & 0 &  &  \\
		& \ddots & \ddots & \\ 
		&   & 1 & 0\\
		\end{pmatrix} $
		
		&
		\newline
		$ \im \asign
		\begin{pmatrix}
		&  &  &  & \li\\
		&  &  & -\li & \\
		&  & \iddots & & \\
		& -\li &  &  & \\
		\li &  &  &  & \\
		\end{pmatrix} $
		
		&
		$\im \asign
		\begin{pmatrix}
		&  &  &  & -\li\\
		&  &  & \li & \\
		&  & \iddots &  & \\
		& \li &  &  & \\
		-\li &  &  &  & \\
		\end{pmatrix}$
		&
		$D$ (odd)
		\\
		\hline
	\end{tabular}
\end{table*}

To construct the normal form of the equation-of-motion matrix $\KNquad$, as given in \eqnref{KN}, one requires to find all the gGEVs of $\Kquad$ in the symplectic orthonormalized form $\evec_{ij}$ (see \appref{App:gsON}), their ranks $D_{ij}$, and classify them within the six different cases defined in \secref{GEVs}. 

The normal form of a generic quadratic quantum Hamiltonian, $\Hop = (1/2)\Rhopvec^T N \Rhopvec$, can be readily obtained as $\Nquad=-J \KNquad$, which reads
\be \label{eq:HNF}
\begin{split}
\Hop& = \sum_{\mathcal{R}_i} \sum_{j=1}^{m_i} \Hop^{(1)}(\lambda_i,D_{ij})+\sum_{\mathcal{C}_i} \sum_{j=1}^{m_i} \Hop^{(2)}(\lambda_i,D_{ij})\\
&+ \sum_{j=1}^{\lzz} \Hop^{(3)}(0,D_{0j}) + \sum_{j=1}^{\nzz} \Hop^{(4)}(0,D_{0j}) \\
& + \sum_{\mathcal{I}_i} \spare{ \sum_{j=1}^{\lii} \Hop^{(5)}(\lambda_i,D_{ij})+\sum_{j=1}^{\nii} \Hop^{(6)}(\lambda_i,D_{ij})}.
\end{split}
\ee 
The six possible types of quadratic Hamiltonians $\Hop^{(\mathfrak{c})}(\lambda,D)$ for $\mathfrak{c}=1,\dots,6$ are given, in quadrature representation, in \tabref{TAB:HamiltonianNF}. 
\begin{table*}
\begin{tabular}{|c|l|}
\hline
$\mathfrak{c}$ & $\Hop^{(\mathfrak{c})}(\lambda,D)=\Rhopvec^T N^{(\mathfrak{c})} \Rhopvec/2$\\
\hline
$1$ & $\Hop^{(1)} (\lambda,D) = \lambda \sum_{k=1}^D \Xop_k\Pop_k+\sum_{k=1}^{D-1}\Xop_k\Pop_{k+1}$\rule{0pt}{4.5ex}\rule[-3.0ex]{0pt}{0pt}\\
\hline
$2$ & $\Hop^{(2)}(\lr + \im \li,D) = \lr \sum_{k=1}^{2D} \Xop_k\Pop_k
        +\li \sum_{k=1}^D\pare{\Xop_{2k}\Pop_{2k-1}-\Xop_{2k-1}\Pop_{2k}} +\sum_{k=1}^{2D-2}\Xop_k\Pop_{k+2}$\rule{0pt}{4.5ex}\rule[-3.0ex]{0pt}{0pt}\\
\hline
$3$ & $\Hop^{(3)}(0,D \; \text{even}) = \asign\sum_{k=1}^{D/2-1}\Xop_k\Pop_{k+1} + (-1)^{D/2+1}(\asign/2)\Xop_{D/2}^2$\rule{0pt}{4.5ex}\rule[-3.0ex]{0pt}{0pt}\\
\hline
$4$ & $\Hop^{(4)}(0,D \; \text{odd})=\sum_{k=1}^{D-1}\Xop_k\Pop_{k+1}$\rule{0pt}{4.5ex}\rule[-3.0ex]{0pt}{0pt}\\
\hline
$5$ & $\Hop^{(5)}(\im \li,D \; \text{even}) =(\asign/2) \sum_{k=1}^{D-1}(-1)^{k+1}\pare{\Xop_k\Xop_{D-k}+\Pop_{k+1}\Pop_{D+1-k}}+ \asign (\nu/2)\sum_{k=1}^D\pare{\Xop_k\Xop_{D+1-k}+\Pop_{k}\Pop_{D+1-k}}$ \rule{0pt}{4.5ex}\rule[-3.0ex]{0pt}{0pt}\\
\hline
$6$ & $\Hop^{(6)}(\im \li,D \; \text{odd})= \im\asign (\nu/2)\sum_{k=1}^D (-1)^{k+1}\pare{\Pop_k\Pop_{D-k+1}+\Xop_k\Xop_{D-k+1}}+\sum_{k=1}^{D-1} \Xop_k\Pop_{k+1}$\rule{0pt}{4.5ex}\rule[-3.0ex]{0pt}{0pt}\\
\hline
\end{tabular}
\caption{Normal form of a quadratic quantum Hamiltonian in the quadrature representation. We denote $\lambda = \lr + \im \li$ (with  $\lr,\li \in \mathbb{R}$) and $\asign=\alpha_{\lambda}(\evec, \tilde \evec)$, where $\evec$ is the generating generalized eigenvector (gGEV) associated to each case. If a generic normal form consists of multiple cases, then the summation in each case is over different modes.
}
\label{TAB:HamiltonianNF}
\end{table*}
Note that there are actually nine different Hamiltonians since $\Hop^{(3)}(0,D)$ and $\Hop^{(5)}(\lambda,D)$ can be either with $\sigma=+1$ or $\sigma=-1$, and  $\Hop^{(6)}(\lambda,D)$ can be either with $\im \sigma=1$ or $\im \sigma=-1$ (here $\asign=\alpha_{\lambda}(\evec, \tilde \evec)$, where $\evec$ is the gGEV associated to each case). These  nine Hamiltonians are canonically inequivalent~\citep{Arnold,Broadbridge1979,Williamson1936}, namely they are not related by a canonical transformation matrix. We emphasize that in \eqnref{eq:HNF} one should notice that every term $\Hop^{(\mathfrak{c})}(\lambda_i,D_{ij})$ acts on different modes such that the total Hamiltonian acts on the $N$ modes of the system.

As can be seen from \tabref{TAB:HamiltonianNF}, a quadratic quantum Hamiltonian in the normal form can have the following terms:
\begin{itemize}
    \item Independent harmonic oscillator (in $\mathfrak{c}=6$)
    \be 
\Xop_k^2  + \Pop_k^2 = 2\bdop_k \bop_k + \id,
\ee  
\item Independent free particle, either in the form (in $\mathfrak{c}=3,5$)
 \be 
 \Xop_k^2 = \frac{1}{2} \pare{2 \bdop_k \bop_k + \bop_k^2+\bop_k^{\dag 2}  + \id},
 \ee 
 or in the form (in $\mathfrak{c}=5$)
  \be 
 \Pop_k^2 = \frac{1}{2} \pare{2 \bdop_k \bop_k - \bop_k^2-\bop_k^{\dag 2}  + \id},
 \ee 
 
    \item Single mode squeezing (in $\mathfrak{c}=1,2$)
    \be 
        \Xop_k\Pop_{k} = \frac{\im}{2} \pare{\bop_k^{ \dag2 } - \bop_k^2},
    \ee 
    \item Two-mode beam-splitter interaction, either in the form (in $\mathfrak{c}=2$)
    \be 
\Xop_k\Pop_{l}  - \Pop_k\Xop_{l} = \im \pare{\bop_k\bdop_l - \bdop_k \bop_l},
\ee  
or in the form (in $\mathfrak{c}=5,6$)
\be 
\Xop_k\Xop_{l}  + \Pop_k\Pop_{l} = \bdop_k \bop_l + \bop_k \bdop_l,
\ee  
\item Two-mode beam-splitter and two-mode squeezing interaction (in $\mathfrak{c}=1,2,3,4$)
    \be 
\Xop_k\Pop_{l} = \frac{\im}{2} \pare{\bop_k\bdop_l - \bdop_k \bop_l} + \frac{\im}{2} \pare{\bdop_k\bdop_l - \bop_k \bop_l}.
\ee  
\end{itemize}

The normal form of a quadratic quantum Hamiltonian greatly simplifies whenever the equation-of-motion matrix $\Kquad$ is diagonalizable, namely $a_i=m_i$  and $D_{ij}=1$ for $\forall i,j$. In that case all the eigenvectors are gGEVs, namely $\evec^{(1)}_{ij}=\evec_{ij}$. The Hamiltonian \eqnref{eq:HNF} then simplifies to
\be \label{eq:HDiagon}
\begin{split}
\Hop =& \sum_{\mathcal{R}_i} \sum_{j=1}^{a_i} \Hop^{(1)}(\lambda_i,1)+\sum_{\mathcal{C}_i} \sum_{j=1}^{a_i} \Hop^{(2)}(\lambda_i,1)\\
& +\sum_{j=1}^{a_0/2} \Hop^{(4)}(0,1) + \sum_{\mathcal{I}_i} \sum_{j=1}^{a_i}\Hop^{(6)}(\lambda_i,1).
\end{split}
\ee 
(Here, the $\sum_{j=1}^{a_i} \Hop^{(\mathfrak{c})}(\lambda_i,1)$ is not a trivial sum since every term acts on a different mode $\Xop_j$, $\Pop_j$).
In \eqnref{eq:HDiagon} there are only four types of Hamiltonian ($\mathfrak{c}=1,2,4,6$) since $D=1$ is odd. Furthermore, note that for the zero eigenvalue  $\Hop^{(4)}(0,1)=0$, which  corresponds to $a_0/2$ zero frequency modes~\citep{Colpa1986a,Colpa1986b}. Notice that \eqnref{eq:HDiagon} is, in general, still a dynamically unstable Hamiltonian since only the last term of \eqnref{eq:HDiagon} represents a set of non-interacting harmonic oscillators. Hence, as mentioned before, the fact that the equation-of-motion matrix $\Kquad$ is a diagonalizable matrix is a necessary but not sufficient condition for dynamical stability. Dynamical stability further requires that  the equation-of-motion matrix $\Kquad$ only has imaginary eigenvalues such that the Hamiltonian \eqnref{eq:HDiagon} further simplifies to a sum of $N$ independent harmonic oscillators 
\be \label{eq:Hstable}
\begin{split}
\Hop &= \sum_{\mathcal{I}_i} \sum_{j=1}^{a_i}\Hop^{(6)}(\lambda_i,1) \\
& = \sum_{\mathcal{I}_i}  \Im(\lambda_i)  \sum_{j=1}^{a_i}  \frac{(\im\asign_{ij})}{2} \pare{\Xop_{ij}^2+\Pop_{ij}^2},
\end{split}
\ee 
where $\asign_{ij} = \alpha_{\lambda_i} (\evec_{ij}, \bar \evec_{ij}) =\evec_{ij}^T J \bar \evec_{ij} = \pm \im$. The double subindex in the quadrature operators used in \eqnref{eq:Hstable} simply denotes the different modes of the system.

In the next \secref{sec:NF_Transf}, we provide step-by-step instructions to construct a real canonical transformation that brings any quadratic quantum Hamiltonian in the normal form defined in this section, namely \eqnref{eq:HNF}. The Bogoliubov transformation~\citep{Bogoljubov1958,Valatin1958,Maldonado1993,Colpa1978} will be thus only a particular case that leads to \eqnref{eq:Hstable} whenever the equation-of-motion matrix $\Kquad$ is diagonalizable and has only imaginary eigenvalues.

\section{Normal Form Transformation}\label{sec:NF_Transf}

In this section we provide instructions to construct a real canonical transformation $T \in \mathbb{R}^{2N\times 2N} $ such that it transforms a generic equation-of-motion matrix, via \eqnref{similarity}, in the corresponding real Jordan normal form $\KNquad$ as defined in \eqnref{KN}. In the following, concise instructions are given and their mathematical background is left for the more specialized literature~\cite{Meyer,Laub1974}.

To give the instructions, it is convenient to make some definitions. The columns of the matrix $T$ will be denoted as
\be \label{Ttotal}
T= \pare{{\bf t}_1 \ldots {\bf t}_n \ldots {\bf t}_{2N}} = \pare{T_+\, T_-},
\ee
where $\dim \pare{{\bf t}_n}= 2N \times 1$ for $n=1, \ldots, 2N$ and $\dim \pare{ T_{\pm}}=2N \times N$.
The matrix $T$ represents a real canonical transformation according to the definition in \secref{sec:QH} if the vectors ${\bf t}_m $ are real and fulfill ${\bf t}_m^T J{\bf t}_n=J_{m n}$, for $m,n=1,\dots,2N$. The two rectangular matrices $T_{\pm}$ are further split into six rectangular matrices as 
\be
T_{\pm} = \pare{T_{\pm}^{(1)}T_{\pm}^{(2)}T_{\pm}^{(3)}T_{\pm}^{(4)}T_{\pm}^{(5)}T_{\pm}^{(6)}},
\ee
where $\dim \pares{ T_{\pm}^{(\mathfrak{c})}}=2N \times N_\mathfrak{c}$ for $\mathfrak{c}=1,\ldots, 6$, with $N_\mathfrak{c}$ to be defined below. As the notation indicates, each $T_{\pm}^{(\mathfrak{c})}$ corresponds to one of the different six cases introduced in \secref{sec:SpecProp}, and one has that $N=\sum_{\mathfrak{c}=1}^6 N_\mathfrak{c}$. In the following we show how to construct $T_{\pm}^{(\mathfrak{c})}$ for every case.

\subsection{Transformation for real pair \texorpdfstring{$\mathfrak{c=1}$}{c=1}}\label{sec:NF_Transf_real}
In this case, $T_{\pm}^{(1)}$, of dimension $2N \times N_1$, is further split into 
\be
T_{\pm}^{(1)} = \pare{T_{1 \pm}^{(1)} \ldots T_{i \pm}^{(1)} \ldots T_{N_\mathcal{R} \pm}^{(1)} },
\ee
where there is a $T_{i \pm}^{(1)}$ for every real pair $\mathcal{R}_i = \cpare{\lambda_i,-\lambda_i}$ ($i=1, \ldots, N_\mathcal{R}$) of dimension  $2N \times a_i$. Let us assume that for every real pair $\mathcal{R}_i$ we have already obtained the gGEVs (in the symplectric orthonormalized form) $\evec_{ij} \in \mathcal{H} (\lambda_i)$ and $\tilde \evec_{ij} \in \mathcal{H} (-\lambda_i)$ and their corresponding ranks $D_{ij}$, for $j=1, \ldots, m_i$. Then, the matrices $T_{i \pm}^{(1)}$ are given  by 
\begin{equation} \label{Tc1}
\begin{aligned}
&\Tquad_{i+}^{(1)}=\pare{ \uvec_{i1}^{(1)} \dots \uvec_{i1}^{(D_{i1})}\dots  \uvec_{im_i}^{(1)} \dots \uvec_{im_i}^{(D_{im_i})}},\\
&\Tquad_{i-}^{(1)}=\pare{ \vvec_{i1}^{(1)} \dots \vvec_{i1}^{(D_{i1})}\dots  \vvec_{im_i}^{(1)} \dots \vvec_{im_i}^{(D_{im_i})}},\\
\end{aligned}
\end{equation}
where
\begin{equation}\label{tsc1}
\begin{aligned}
\uvec_{ij}^{(k)}&= (\Kquad-\lambda_i \mathbb{1})^{k-1} \evec_{ij},\\
\vvec_{ij}^{(k)}&= (-1)^{D_{ij}-k}(\Kquad+\lambda_i \mathbb{1})^{D_{ij}-k} \tilde{\evec}_{ij},
\end{aligned}
\end{equation}
for $k=1,...,D_{ij}$. Note that $N_1 = \sum_{i=1}^{N_\mathcal{R}} a_i$.

\subsection{Transformation for complex quadruplet \texorpdfstring{$\mathfrak{c=2}$}{c=2}}\label{sec:NF_Transf_complex}
In this case, $T_{\pm}^{(2)}$, of dimension $2N \times N_2$, is further split into 
\be
T_{\pm}^{(2)} = \pare{T_{1 \pm}^{(2)} \ldots T_{i \pm}^{(2)} \ldots T_{N_\mathcal{C} \pm}^{(2)} },
\ee
where there is a $T_{i \pm}^{(2)}$ for every complex quadruplet $\mathcal{C}_i = \cpare{\lambda_i, - \lambda_i, \bar \lambda_i, - \bar \lambda_i}$ ($i=1, \ldots, N_\mathcal{C}$) of dimension  $2N \times 2 a_i$.  
Let us assume that for every complex quadruplet we have already obtained the gGEVs (in the symplectric orthonormalized form) $\evec_{ij} \in \mathcal{H} (\lambda_i)$ and $\tilde \evec_{ij} \in \mathcal{H} (-\lambda_i)$ and their corresponding ranks $D_{ij}$, for $j=1, \ldots, m_i$. Then, the matrices $T_{i \pm}^{(2)}$ are given  by 
\begin{equation} 
\begin{aligned}
&\Tquad_{i+}^{(2)}= \pare{ \uvec_{i1}^{(1)} \dots \uvec_{i1}^{(2D_{i1})}\dots  \uvec_{im_i}^{(1)} \dots \uvec_{im_i}^{(2D_{im_i})} },\\
&\Tquad_{i-}^{(2)}=\pare{ \vvec_{i1}^{(1)} \dots \vvec_{i1}^{(2D_{i1})}\dots  \vvec_{im_i}^{(1)} \dots \vvec_{im_i}^{(2D_{im_i})}},\\
\end{aligned}
\end{equation}
where
\begin{equation} 
\begin{aligned}
\uvec_{ij}^{(k)} &= 
\begin{cases}
\sqrt{2} \mathrm{Re} \left(\zvec_{ij}^{((k+1)/2)}\right)  & k=\text{odd},\\
\sqrt{2} \mathrm{Im} \left(\zvec_{ij}^{(k/2)}\right) & k=\text{even},\\
\end{cases}\\
\vvec_{ij}^{(k)} &= 
\begin{cases}
\sqrt{2} \mathrm{Re} \left(\wvec_{ij}^{((k+1)/2)}\right)  & k=\text{odd},\\
-\sqrt{2} \mathrm{Im} \left(\wvec_{ij}^{(k/2)}\right) & k=\text{even},\\
\end{cases}
\end{aligned}
\end{equation}
with
\begin{equation}\label{}
\begin{aligned}
\zvec_{ij}^{(k)}&= (\Kquad-\lambda_i \mathbb{1})^{k-1} \evec_{ij}\\
\wvec_{ij}^{(k)}&= (-1)^{D_{ij}-k}(\Kquad+\lambda_i \mathbb{1})^{D_{ij}-k} \tilde{\evec}_{ij},
\end{aligned}
\end{equation}
for $k=1,...,D_{ij}$. Note that $N_2 = 2 \sum_{i=1}^{N_\mathcal{C}} a_i$.

\subsection{Transformation for zero eigenvalues \texorpdfstring{$\mathfrak{c=3}$}{c=3}}\label{sec:NF_Transf_zero-even}
In this case, $T_{\pm}^{(3)}$, of dimension $2N \times N_3$, does not need to be further split. Let us assume that we have already obtained $\lzz$ gGEVs (in the symplectric orthonormalized form) $\evec_{0j} \in \mathcal{H} (0) $, their even ranks $D_{0j}$ and the values $\sigma_{0j} = \alpha_0(\evec_{0j},\evec_{0j})$ for $j=1, \ldots,  \lzz$. Then, the matrices $T_{\pm}^{(3)}$ are given  by~\cite{Meyer,Laub1974}
\begin{equation} 
\begin{aligned}
&\Tquad_{+}^{(3)}=\pare{ \uvec_{01}^{(1)} \dots \uvec_{01}^{(D_{01}/2)}\dots  \uvec_{0 \lzz}^{(1)} \dots \uvec_{0 \lzz}^{(D_{0 \lzz}/2)}},\\
&\Tquad_{-}^{(3)}= \pare{ \vvec_{01}^{(1)} \dots \vvec_{01}^{(D_{01}/2)}\dots  \vvec_{0\lzz}^{(1)} \dots \vvec_{0 \lzz}^{(D_{0 \lzz}/2)}},\\
\end{aligned}
\end{equation}
where
\begin{equation}\label{}
\begin{aligned}
\uvec_{0j}^{(k)}&= \sigma_{0j}^{k-1} K^{k-1} \evec_{0j},\\
\vvec_{ij}^{(k)}&= (-\sigma_{0j})^{D_{0j}-k} K^{D_{0j}-k} \evec_{0j},
\end{aligned}
\end{equation}
for $k=1,...,D_{0j}/2$. Note that $N_3 =  \sum_{j=1}^{\lzz} D_{0j}/2$.

\subsection{Transformation for zero eigenvalues \texorpdfstring{$\mathfrak{c=4}$}{c=4}}\label{sec:NF_Transf_zero-odd}
Compared to the other cases, case $\mathfrak{c=4}$ requires an additional step~\cite{Meyer,Laub1974}. As explained in \appref{App:gsON-c4}, the set of gGEVs (after having performed the generalized symplectic orthonormalization) $\lbrace \evec_{01}, \dots, \evec_{0 2\nzz} \rbrace$ of odd rank $D_{0j}$ is transformed to a new set of gGEVs $\lbrace \fvec_{01}, \dots, \fvec_{0 \nzz}, \hvec_{01}, \dots, \hvec_{0 \nzz} \rbrace$, which satisfy $\alpha_0(\fvec_{0j},\hvec_{0j'})=\delta_{jj'}$ and $\alpha_0(\fvec_{0j},\fvec_{0j'})=\alpha_0(\hvec_{0j},\hvec_{0j'})=0$ for $j, j'=1,\dots, \nzz$. After this additional step has been done, one can proceed.

$T_{\pm}^{(4)}$, of dimension $2N \times N_4$, can be obtained from the $2\nzz$ gGEVs $\lbrace \fvec_{01}, \dots, \fvec_{0 \nzz}, \hvec_{01}, \dots, \hvec_{0 \nzz} \rbrace$ by~\cite{Meyer,Laub1974}
\begin{equation} \label{Tc4}
\begin{aligned}
&\Tquad_{+}^{(4)}=\lbrace \uvec_{01}^{(1)}, \dots, \uvec_{01}^{(D_{01})},\dots,  \uvec_{0 \nzz}^{(1)}, \dots, \uvec_{0 \nzz}^{(D_{0 \nzz})}\rbrace,\\
&\Tquad_{-}^{(4)}=\lbrace \vvec_{01}^{(1)}, \dots, \vvec_{01}^{(D_{01})},\dots,  \vvec_{0\nzz}^{(1)}, \dots, \vvec_{0 \nzz}^{(D_{0 \nzz})}\rbrace,\\
\end{aligned}
\end{equation}
where   
\begin{equation}\label{tsc4}
\begin{aligned}
\uvec_{0j}^{(k)}&= K^{k-1} \fvec_{0j},\\
\vvec_{ij}^{(k)}&= (-1)^{D_{0j}-k} K^{D_{0j}-k} \hvec_{0j},
\end{aligned}
\end{equation}
for $k=1,...,D_{ij}$. Note that $N_4 = \sum_{j=1}^{\nzz} D_{0j} $ and $N_3+N_4 = a_0/2$.

\subsection{Transformation for imaginary pair \texorpdfstring{$\mathfrak{c=5}$}{c=5}}\label{sec:NF_Transf_imag-even}
In this case, $T_{\pm}^{(5)}$, of dimension $2N \times N_5$, is further split into 
\be
T_{\pm}^{(5)} = \pare{T_{1 \pm}^{(5)} \ldots T_{i \pm}^{(5)} \ldots T_{L_\mathcal{I} \pm}^{(5)} },
\ee
where there is a $T_{i \pm}^{(5)}$ for every imaginary pair $\mathcal{I}_i = \cpare{\lambda_i,-\lambda_i}$ ($i=1, \ldots, L_\mathcal{I}$) of dimension  $2N \times a_i^{(5)}$, where $a_i^{(5)} = \sum_{j=1}^{\lii} D_{ij}$. $L_\mathcal{I}$ is the number of distinct imaginary pairs with the gGEVs belonging to the case $\mathfrak{c}=5$. Let us assume that for every imaginary pair $\mathcal{I}_i$ we have already obtained the gGEVs (in the symplectric orthonormalized form) $\evec_{ij} \in \mathcal{H} (\lambda_i)$, $\bar \evec_{ij} \in \mathcal{H} (\bar\lambda_i)$, their corresponding even ranks $D_{ij}$ and the values $\sigma_{ij} = \alpha_i(\evec_{ij},\bar\evec_{ij})$ for $j=1, \ldots,  \lii$. Then, the matrices $T_{i \pm}^{(5)}$ are given  by~\cite{Meyer,Laub1974}
\begin{equation} 
\begin{aligned}
&\Tquad_{i+}^{(5)}=\lbrace \uvec_{i1}^{(1)}, \dots, \uvec_{i1}^{(D_{i1})},\dots,  \uvec_{i \lii}^{(1)}, \dots, \uvec_{i \lii}^{(D_{i\lii})}\rbrace,\\
&\Tquad_{i-}^{(5)}=\lbrace \vvec_{i1}^{(1)}, \dots, \vvec_{i1}^{(D_{i1})},\dots,  \vvec_{i \lii}^{(1)}, \dots, \vvec_{i \lii}^{(D_{i\lii})}\rbrace,\\
\end{aligned}
\end{equation}
where
\begin{equation}\label{}
\begin{aligned}
\uvec_{ij}^{(k)} &= 
\begin{cases}
\sqrt{2} \mathrm{Re} (\zvec_{ij}^{(k)}) , \quad k=\text{odd},\\
\sqrt{2} \mathrm{Im} (\zvec_{ij}^{(k)}) , \quad k=\text{even},\\
\end{cases}\\
\vvec_{ij}^{(k)} &= 
\begin{cases}
\sqrt{2} \mathrm{Re} (\wvec_{ij}^{(k)}) , \quad k=\text{odd},\\
-\sqrt{2} \mathrm{Im} (\wvec_{ij}^{(k)}) , \quad k=\text{even},\\
\end{cases}
\end{aligned}
\end{equation}
with
\begin{equation}\label{}
\begin{aligned}
\zvec_{ij}^{(k)}&= (\Kquad-\lambda_i \mathbb{1})^{k-1} \evec_{ij},\\
\wvec_{ij}^{(k)}&= \sigma_{ij}(-1)^{k} \bar \zvec_{ij}^{(D_{ij}+1-k)},
\end{aligned}
\end{equation}
for $k=1,...,D_{ij}$. Note that $N_5 = \sum_{i=1}^{N_\mathcal{I}} a_i^{(5)}$.

\subsection{Transformation for imaginary pair \texorpdfstring{$\mathfrak{c=6}$}{c=6}}\label{sec:NF_Transf_imag-odd}
In this case, $T_{\pm}^{(6)}$, of dimension $2N \times N_6$, is further split into 
\be
T_{\pm}^{(6)} = \pare{T_{1 \pm}^{(6)} \ldots T_{i \pm}^{(6)} \ldots T_{S_\mathcal{I} \pm}^{(6)} },
\ee
where there is a $T_{i \pm}^{(6)}$ for every imaginary pair $\mathcal{I}_i = \cpare{\lambda_i,-\lambda_i}$ ($i=1, \ldots, S_\mathcal{I}$) of dimension  $2N \times a_i^{(6)}$, where $a_i^{(6)} = \sum_{j=1}^{\nii} D_{ij}$. $S_\mathcal{I}$ is the number of distinct imaginary pairs with the gGEVs belonging to the case $\mathfrak{c}=6$. We remark that in general, an imaginary pair $\mathcal{I}_i$ can have gGEVs of both cases $\mathfrak{c}=5,6$, such that $L_\mathcal{I}+S_\mathcal{I} \geq N_\mathcal{I}$. Let us assume that for every imaginary pair $\mathcal{I}_i$ we have already obtained the gGEVs (in the symplectric orthonormalized form) $\evec_{ij} \in \mathcal{H} (\lambda_i)$, $\bar \evec_{ij} \in \mathcal{H} (\bar\lambda_i)$, their corresponding odd ranks $D_{ij}$ and the values $\sigma_{ij} = \alpha_i(\evec_{ij},\bar\evec_{ij})$ for $j=1, \ldots,  \nii$. Then, the matrices $T_{i \pm}^{(6)}$ are given  by~\cite{Meyer,Laub1974}
\begin{equation} \label{Tc6}
\begin{aligned}
&\Tquad_{i+}^{(6)}=\lbrace \uvec_{i1}^{(1)}, \dots, \uvec_{i1}^{(D_{i1})},\dots,  \uvec_{i \nii}^{(1)}, \dots, \uvec_{i \nii}^{(D_{i\nii})}\rbrace,\\
&\Tquad_{i-}^{(6)}=\lbrace \vvec_{i1}^{(1)}, \dots, \vvec_{i1}^{(D_{i1})},\dots,  \vvec_{i \nii}^{(1)}, \dots, \vvec_{i \nii}^{(D_{i\nii})}\rbrace,\\
\end{aligned}
\end{equation}
where
\begin{equation} \label{tsc6}
    \begin{aligned}
    \uvec_{ij}^{(k)} &= \sqrt{2} \mathrm{Re} (\zvec_{ij}^{(k)})\\
    \vvec_{ij}^{(k)} &= \sqrt{2} \Im (\bar\wvec_{ij}^{(k)}),
    \end{aligned}
\end{equation}
with
\begin{equation}\label{}
\begin{aligned}
\zvec_{ij}^{(k)}&= (\Kquad-\lambda_i \mathbb{1})^{k-1} \evec_{ij},\\
\wvec_{ij}^{(k)}&= \sigma_{ij}(-1)^{k} \bar \zvec_{ij}^{(D_{ij}+1-k)},
\end{aligned}
\end{equation}
for $k=1,...,D_{ij}$. Note that $N_6 = \sum_{i=1}^{S_\mathcal{I}} a_i^{(6)}$ and $a_i^{(5)}+a_i^{(6)}=a_i$.

\section{Examples}\label{sec:Examples}

To illustrate the tools presented so far, in this section we provide two examples. In \secref{ssec:2HO} we discuss the stability diagram of two quantum harmonic oscillators with a quadratic position coupling. The different normal forms throughout the stability diagram are given. In \secref{ssec:detailed}, we provide an example of  a particular quadratic Hamiltonian of four harmonic oscillators and we give details on the derivation of the canonical transformation that brings it into its normal form, following the instructions given in \secref{sec:NF_Transf}.

\subsection{Stability diagram of the two-mode Hamiltonian with position coupling}\label{ssec:2HO}
Let us consider the following standard two-mode quadratic Hamiltonian
\be\label{2HO-Hamiltonian}
\Hop = \frac{1}{2} \left( \xop_1^2 + \pop_1^2  \right) + \frac{\freq}{2} \left( \xop_2^2 + \pop_2^2  \right) + \coup \xop_1 \xop_2,
\ee
which depends on the dimensionless real parameters $\freq$ and $\coup$.  The $M$ matrix (recall \eqnref{QH}) is given by
\begin{equation}\label{2HO-M}
\begin{aligned}
M = \begin{pmatrix}
    1 & \coup  & 0 & 0 \\
    \coup  & \freq & 0 & 0 \\
    0 & 0 & 1 & 0 \\
    0 & 0 & 0 & \freq \\
    \end{pmatrix}.
\end{aligned}
\end{equation}
and hence the equation-of-motion matrix $K$ by
\begin{equation}\label{2HO-K}
\begin{aligned}
K = JM =
    \begin{pmatrix}
    0 & 0 & 1 & 0 \\
    0 & 0 & 0 & \freq  \\
    -1 & -\coup  & 0 & 0 \\
    -\coup  & -\freq  & 0 & 0 \\
    \end{pmatrix}.
\end{aligned}
\end{equation}

Depending on the values of $(\freq,\coup)$ one can encounter different normal forms, see \figref{fig:StabilityDiagram}:
\begin{figure}[t]
	\centering
	\includegraphics[width=0.8\linewidth]{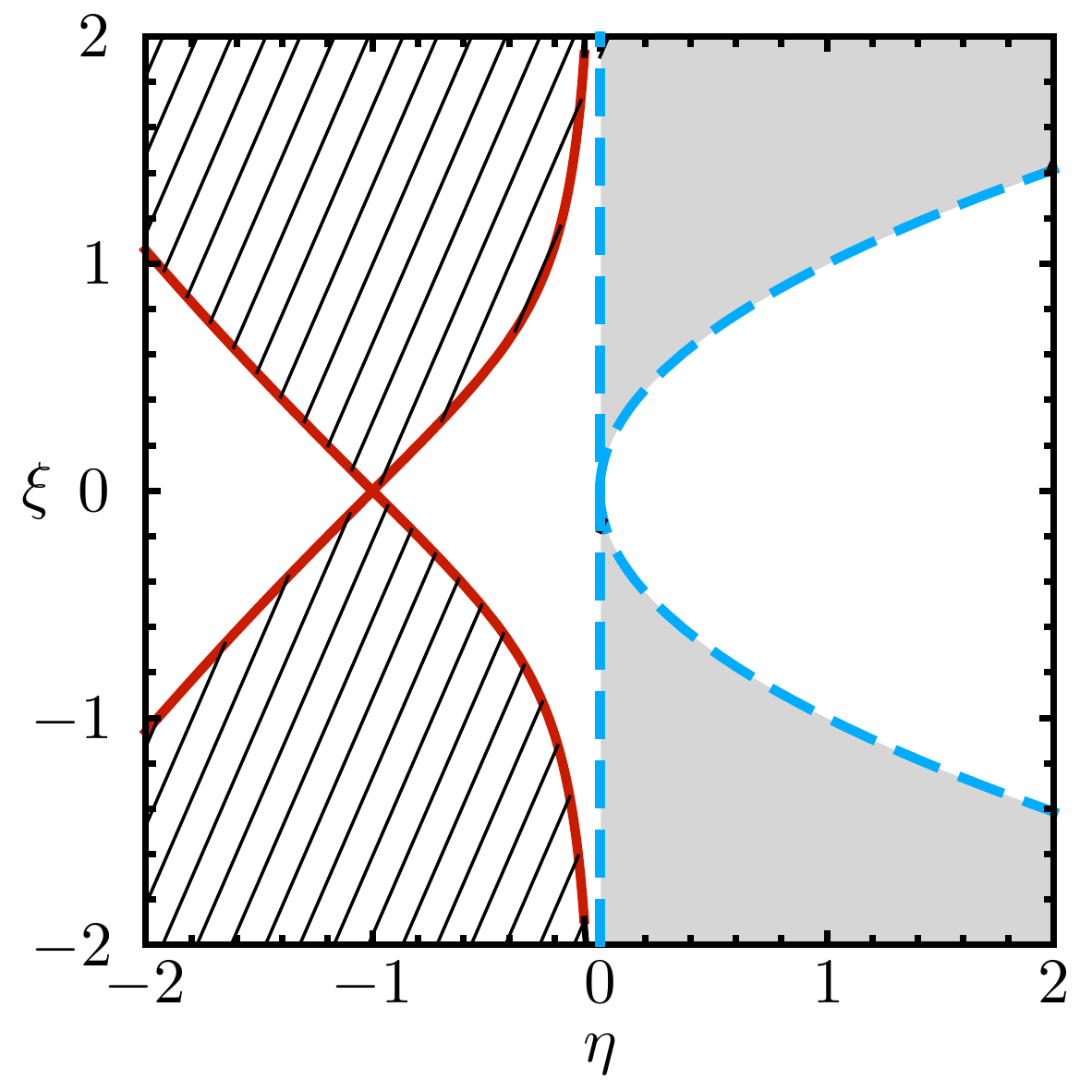}
	\caption{Spectrum of $\Kquad$ as function of dimensionless parameters $\freq$ and $\coup$. White area: two imaginary pairs, diagonalizable. Hatched area: a complex quadruplet, diagonalizable. Gray area: an imaginary pair and a real pair, diagonalizable. Red line: degenerate imaginary pair, nondiagonalizable. Blue dashed line: an imaginary pair and zero eigenvalue, nondiagonalizable.}
	\label{fig:StabilityDiagram}
\end{figure}
\begin{itemize}
    \item \underline{White area}: one has two non-degenerate imaginary pairs $\mathcal{I}_1 = \lbrace \lambda_1, \bar \lambda_1 \rbrace $ and $\mathcal{I}_2=\lbrace \lambda_2, \bar \lambda_2 \rbrace$, with $a_i=m_i=D_i=1$ for $i=1,2$. Since $D_i$ is odd, both pairs correspond to $\mathfrak{c}=6$. $\sigma_{ij}$ depends on the value of $\freq$. For arbitrary gGEVs at any point, one has: if $\freq > 0$,  $\sigma_{i1}=\im \mathrm{sgn}[\alpha_{\lambda_i}(\gvec_{i1},\bar\gvec_{i1})/\im]=-\im$, and if $\freq <0$, $\sigma_{11}=\im \mathrm{sgn}[\alpha_{\lambda_1}(\gvec_{11},\bar\gvec_{11})/\im]=-\im$ and $\sigma_{21}=\im \mathrm{sgn}[\alpha_{\lambda_2}(\gvec_{21},\bar\gvec_{21})/\im]=\im$. The normal form Hamiltonian is then
    \begin{equation}\label{}
    \begin{aligned}
    \Hop = &\im \sigma_{11} \frac{\Im(\lambda_1)}{2} \left( \hat{X}_1^2+\hat{P}_1^2 \right)\\
    &+  \im \sigma_{21}\frac{\Im(\lambda_2)}{2} \left( \hat{X}_2^2+\hat{P}_2^2 \right),
    \end{aligned}
    \end{equation}
    which represents to two independent harmonic oscillators. Hence, the white area is the region where the system is dynamically stable.
    \item \underline{Gray area}: one has one non-degenerate imaginary pair $\mathcal{I}_1 = \lbrace \lambda_1, \bar \lambda_1 \rbrace$ and one real pair $\mathcal{R}_2= \lbrace \lambda_2, - \lambda_2 \rbrace$, with $a_i=m_i=D_i=1$ for $i=1,2$. Since $D_1$ is odd, $\mathcal{I}_1$ corresponds to case $\mathfrak{c}=6$. For an arbitrary gGEV at any point, $\sigma_{11}=\im \mathrm{sgn}[\alpha_{\lambda_1}(\gvec_{11},\bar\gvec_{11})]=-\im$. The real pair $\mathcal{R}_2$ corresponds to  $\mathfrak{c}=1$. Hence, the normal form Hamiltonian is
    \begin{equation}\label{}
    \begin{aligned}
    \Hop = \frac{\Im(\lambda_1)}{2} \left( \hat{X}_1^2+\hat{P}_1^2 \right)  +  \lambda_2 \Xop_2 \Pop_2,
    \end{aligned}
    \end{equation}
    which represents  two uncoupled modes, one being squeezed. The single-mode squeezing term induces dynamical instability.
    \item \underline{White hatched area}: one has one non-degenerate complex quadruplet $\mathcal{C} = \lbrace \lambda, -\lambda,\bar \lambda, -\bar \lambda \rbrace$ with $a=m=D=1$ of case $\mathfrak{c}=2$. The normal form of the quadratic Hamiltonian is thus
    \begin{equation}\label{}
    \begin{aligned}
    \Hop = &\Re(\lambda) \left( \Xop_1 \Pop_1 + \Xop_2 \Pop_2 \right) \\& + \Im(\lambda) \left( \Xop_2 \Pop_1 - \Xop_1 \Pop_2 \right),
    \end{aligned}
    \end{equation}
    which corresponds to two modes that are independently squeezed and interact via a beam-splitter interaction. The single-mode squeezing terms induce dynamical instability.
    \item \underline{Dashed blue line} (except the point $\freq=\coup = 0$, see below): one has a non-degenerate imaginary pair  $\mathcal{I}_1 = \lbrace \lambda_1, \bar\lambda_1\rbrace$ with $a_1=m_1=D_1=1$ and $\sigma_{11}=-\im$, which corresponds to case $\mathfrak{c}=6$, and one zero eigenvalue $\lambda_0=0$ with $a_0=2$, $m_0=1$, and hence $D_0=2$, which corresponds to case $\mathfrak{c}=3$. $\sigma_{01}$ depends on the value of $\freq$. For an arbitrary gGEV at any point, one has: if $\freq > 0$ ($\freq = 0$) then  $\sigma_{01}=\mathrm{sgn}[\alpha_{0}(\gvec_{01},\gvec_{01})]=1(-1)$. The  normal form of the quadratic Hamiltonian is given by
    \begin{equation}\label{}
    \begin{aligned}
    \Hop = \frac{\Im(\lambda_1)}{2} \left( \Xop_1 ^2 + \Pop_1^2 \right) + \frac{\sigma_{01}}{2} \Xop_2^2.
    \end{aligned}
    \end{equation}
    This Hamiltonian corresponds to a harmonic oscillator and an uncoupled free particle, which induces dynamical instability. The dashed blue lines correspond to $\freq=0$ and $\coup=\pm \sqrt{ \freq}$ for $\freq > 0$.
    \item \underline{Solid red line} (except the point $\freq=-1, \coup = 0$, see below): one has a degenerate imaginary pair $\mathcal{I}_1 = \lbrace \lambda_1, \bar\lambda_1\rbrace$ with $a_1=2$, $m_1=1$, and $D_1=2$. Since $D_1=2$ is even, this corresponds to case $\mathfrak{c}=5$. $\sigma_{11}$ depends on the value of $\freq$. For an arbitrary gGEV at any point, one has: if $\freq > -1$ ($\freq < -1 $) then $\sigma_{11}=\im \mathrm{sgn}[\alpha_{0}(\gvec_{01},\gvec_{01})/\im]=\im(-\im)$. The normal form of the quadratic Hamiltonian is then given by
    \be 
    \begin{split}
    \Hop =&  \frac{\sigma_{11}}{2}\pare{\Xop_1^2+ \Pop_2^2} \\
    &+\sigma_{11} \frac{\Im (\lambda_1)}{2} \pare{\Xop_1 \Xop_2 + \Pop_1 \Pop_2}.
    \end{split}
    \ee 
    This Hamiltonian corresponds to two free particles coupled by a beam-splitter type interaction. The free particle terms generate dynamical instability. The solid red lines correspond to $\coup=\pm \sqrt{(2\freq^2-\freq^4-1)/(4\freq)}$ for $\eta < 0$.
    \item \underline{Special point $\freq =\coup = 0$}: the matrix $\Mquad$ as given in \eqnref{2HO-M} is trivially in the normal form $\Hop = (1/2) \left( \xop_1^2 + \pop_1^2  \right)$ but recall that the Hamiltonian still describes the dynamics of two modes. The eigenvalues of $\Kquad$ are one non-degenerate imaginary pair  $\mathcal{I}_1 = \lbrace \im, -\im\rbrace$ with $a_1=m_1=1=D_1$ and $\sigma_{11}=-\im$, which corresponds to case $\mathfrak{c}=6$, and one zero eigenvalue $\lambda_0=0$, but now with $a_0=m_0=2$, and hence $D_0=1$, which corresponds to case $\mathfrak{c}=4$. The zero eigenvalues appearing at this point describe a zero frequency mode~\citep{Colpa1986a,Colpa1986b}.
    \item \underline{Special point $\freq=-1, \coup = 0$}: the matrix $\Mquad$ as given in \eqnref{2HO-M} is already in the normal form, which consists of the two independent harmonic oscillators: $\Hop = (1/2) \left( \xop_1^2 + \pop_1^2  \right) - (1/2) \left( \xop_2^2 + \pop_2^2  \right) $. One has a degenerate imaginary pair $\mathcal{I}_1 = \lbrace \im, -\im\rbrace$ but now with $a_1=m_1=2$, and $D_{11}=D_{12}=1$. Since $D_1=1$ is odd, this corresponds to case $\mathfrak{c}=6$. The values of $\sigma_{ij}$ are $\sigma_{11}=-\im$ and $\sigma_{12}=\im$, as can immediately be seen from the signs in the Hamiltonian.
\end{itemize}

With the above list one therefore sees that the simple example of two harmonic oscillators coupled in position covers all nine nonequivalent types of normal forms of a quadratic Hamiltonian.

\subsection{A detailed example}\label{ssec:detailed}

In this subsection, we consider a four-mode quadratic Hamiltonian $\hat{H}=(1/2)\mathbf{\Rop}^T \Mquad \mathbf{\Rop}$, with the matrix $\Mquad$ given by
\be\label{QH-example}
\Mquad = 
\left(
\begin{array}{cccccccc}
 -21 & -11 & -17 & -45 & 16 & 7 & -3 & 22 \\
 -11 & 2 & -6 & -15 & 3 & 6 & -3 & 9 \\
 -17 & -6 & -3 & -29 & 8 & 4 & 0 & 16 \\
 -45 & -15 & -29 & -60 & 19 & 16 & 0 & 33 \\
 16 & 3 & 8 & 19 & -5 & -6 & 0 & -11 \\
 7 & 6 & 4 & 16 & -6 & -1 & 0 & -8 \\
 -3 & -3 & 0 & 0 & 0 & 0 & 3 & 0 \\
 22 & 9 & 16 & 33 & -11 & -8 & 0 & -17 \\
\end{array}
\right)
\ee
and $\Ropvec = (\hat {x}_1, \hat {x}_2, \hat{ x}_3, \hat {x}_4,\hat {p}_1, \hat {p}_2, \hat {p}_3, \hat {p}_4)^T$ the physical modes. The corresponding equation of motion matrix is then obtained as $\Kquad = J \Mquad$, and it has the following eigenvalues:
\begin{itemize}
    \item one degenerate real pair $\mathcal{R}_1 = \lbrace \lambda_1,-\lambda_1 \rbrace =\lbrace2,-2 \rbrace$ with the multiplicities $a_{1}=2$ and $m_1=1$. To the eigenvalue $\lambda_{1}=2$ corresponds a gGEV $\gvec_{1}= (-1,2,0,1,3,-1,1,0)^T$ and to the eigenvalue $-\lambda_{1}=-2$ corresponds a gGEV $\tilde\gvec_{11}= (3,-6,0,-2,-12,8,-3,0)^T$. Both have the rank $D_{11}=2$. This is the case $\mathfrak{c}=1$.
    \item one zero eigenvalue with multiplicities $a_0=2$ and $m_0=2$. The corresponding gGEVs, which are at the same time also eigenvectors, are $\gvec_{01}=(2, 2, -1, 1, 1, 0, 4, 4)^T$ and $\gvec_{02}=(0, 0, 1, 1, 3, 2, 0, 0)^T$. Since their rank is $D_{01}=1$, they are in case $\mathfrak{c}=4$.
    \item one non-degenerate imaginary pair $\mathcal{I}_2 =   \lbrace \lambda_2,-\lambda_2 \rbrace=\lbrace 3 \im,-3 \im \rbrace$ with the multiplicities $a_{2}=1$ and $m_{2}=1$. To the eigenvalue $\lambda_{2}=3 \im$ corresponds a gGEV $\gvec_{21}= (1, 1, -\im, 1, 2 - \im, 1 - \im, 3, 2)^T$ and to the eigenvalue $\bar\lambda_{2}=-3 \im$ corresponds a gGEV $\bar\gvec_{21}= (1, 1, \im, 1, 2 + \im, 1 + \im, 3, 2)^T$. Both have the rank $D_{21}=1$, so this is the case $\mathfrak{c}=6$.
\end{itemize}

Let us start with the real pair. The product \eqnref{Omega} in the vector form has two elements because $D_{11}=2$ and it is $\Omega_{\lambda_1}(\gvec_{11},\tilde\gvec_{11})=-10\mathbb{1} + 13(K-2 \mathbb{1})$. Doing the symplectic orthonormalization as described in \appref{App:gsON-rc} gives two new gGEVs 
\be
\begin{split}
    \evec_{11}&=\frac{-1}{10}(3, 2, 0, 23, 43, 3, 23, 26)^T,\\
    \tilde\evec_{11}&=\frac{-1}{40}(14, 37, 0, 34, 74, -6,51, 65)^T,
\end{split}
\ee
for which $\Omega_{\lambda_1}(\evec_{11},\tilde\evec_{11})=\mathbb{1}$. Moreover, following the instructions given in \eqnref{tsc1}, one obtains that $\uvec^{(1)}_{11}=\evec_{11}$, $\vvec^{(2)}_{11}=\tilde\evec_{11}$ and 
\be
\begin{split}
    \uvec^{(2)}_{11}&=(-2, 0, 0, -2, -2, -2, -2, -4)^T,\\
    \vvec^{(1)}_{11}&=\frac{-1}{2}(2, 1, 0, 2, 2, 2, 3, 5)^T.
\end{split}
\ee
From here, using \eqnref{Tc1} one obtains $T_{1+}= (\uvec^{(1)}_{11}\uvec^{(2)}_{11})$ and $T_{1-}= (\vvec^{(1)}_{11}\vvec^{(2)}_{11})$.

Let us now look at the zero eigenvalue. The rank $D_{01}$ is odd, so this is $\mathfrak{c}=4$. The product \eqnref{Omega} in the vector form for both gGEVs has just a single element because $D_{0}=1$ and this element is simply \eqnref{alpha}. Moreover, since  $\mathfrak{c}=4$, for both gGEVs $\gvec_{01}$ and $\gvec_{02}$ it is $\alpha_0(\gvec_{01},\gvec_{01})=0$ and $\alpha_0(\gvec_{02},\gvec_{02})=0$, but $\alpha_0(\gvec_{01},\gvec_{02})=2$.
Doing the symplectic orthonormalization for $\mathfrak{c}=4$ as described in \appref{App:gsON-c4}  gives the two new gGEVs
\be
\begin{split}
    \fvec_{01}&=(2, 2, -1, 1, 1, 0, 4, 4)^T,\\
    \hvec_{01}&=(0, 0, 1/2, 1/2, 3/2, 1, 0, 0)^T,
\end{split}
\ee
for which $\alpha_0(\fvec_{01},\fvec_{01})=\alpha_0(\hvec_{01},\hvec_{01})=0$, and $\alpha_0(\fvec_{01},\hvec_{01})=1$. $\Tquad_{0\pm}$ as given in \eqnref{Tc4} is then simply $\Tquad_{0+}=(\fvec_{01})$ and $\Tquad_{0-}=(\hvec_{01})$.

Finally, in the case of the imaginary eigenvalue, the rank $D_{21}$ is odd, so this is $\mathfrak{c}=6$. The product \eqnref{Omega} in the vector form again has just a single element because $D_{21}=1$ and this element is simply \eqnref{alpha}, with the value $\alpha_{\lambda_2}(\gvec_{21},\bar\gvec_{21})= -2 \im$. Note that this means that $\asign_{21}=-\im$. Doing the symplectic orthonormalization for $\mathfrak{c}=6$ as described in \appref{App:gsON-c56}  gives us a new gGEV
\be
\evec_{21}=1/\sqrt{2}(1, 1, -\im, 1, 2 - \im, 1 - \im, 3, 2)^T
\ee
with $\alpha_{\lambda_2}(\evec_{21},\bar\evec_{21})= - \im$. $\Tquad_{2\pm}$ as given in \eqnref{Tc6} is then $\Tquad_{2+}=(\sqrt{2}\Re[\evec_{21}])$ and $\Tquad_{2-}=(-\sqrt{2}\Im[\evec_{21}])$.

Constructing the total transformation as in \eqnref{Ttotal}, where $\Tquad_\pm = (\Tquad_{1\pm}\Tquad_{0\pm}\Tquad_{2\pm})$, one obtains the canonical transformation 
\be \label{T-example}
\Tquad = \frac{1}{40}
\left(
\begin{array}{cccccccc}
  -12 & -80 & 80 & 40 & -40 & -14 & 0 & 0 \\
 -80 & 0 & 80 & 40 & -20 & -37 & 0 & 0 \\
 0 & 0 & -40 & 0 & 0 & 0 & 20 & -40 \\
 -92 & -80 & 40 & 40 & -40 & -34 & 20 & 0 \\
 -172 & -80 & 40 & 80 & -40 & -74 & 60 & -40 \\
 -12 & -80 & 0 & 40 & -40 & 6 & 40 & -40 \\
 -92 & -80 & 160 & 120 & -60 & -51 & 0 & 0 \\
 -104 & -160 & 160 & 80 & -100 & -65 & 0 & 0 \\
\end{array}
\right).
\ee

The normal form of the equation-of-motion matrix is $\KNquad=\Tquad^{-1} \Kquad T$, and of the matrix in the Hamiltonian 
\be
\Nquad = \Tquad^T \Mquad T =
\left(
\begin{array}{cccccccc}
 0 & 0 & 0 & 0 & 2 & 1 & 0 & 0 \\
 0 & 0 & 0 & 0 & 0 & 2 & 0 & 0 \\
 0 & 0 & 0 & 0 & 0 & 0 & 0 & 0 \\
 0 & 0 & 0 & 3 & 0 & 0 & 0 & 0 \\
 2 & 0 & 0 & 0 & 0 & 0 & 0 & 0 \\
 1 & 2 & 0 & 0 & 0 & 0 & 0 & 0 \\
 0 & 0 & 0 & 0 & 0 & 0 & 0 & 0 \\
 0 & 0 & 0 & 0 & 0 & 0 & 0 & 3 \\
\end{array}
\right),
\ee
which can be compared with \eqnref{QH-example}.
The normal form of the Hamiltonian \eqnref{QH-example} is
\be
\Hop = 2 (\Xop_1 \Pop_1 + \Xop_2 \Pop_2) + \Xop_1 \Pop_2 + \frac{3}{2} \left(  \Xop_4^2 + \Pop_4^4\right).
\ee
The normal modes $\Rhopvec = (\hat {X}_1, \hat {X}_2, \hat{X}_3, \hat {X}_4,\hat {P}_1, \hat {P}_2, \hat {P}_3, \hat {P}_4)^T$ are a linear combination of the physical modes, and can be obtained using the transformation \eqnref{T-example} as $\Rhopvec = T^{-1} \Ropvec$. Note that the normal mode $(\Xop_3, \Pop_3)$ is a zero-frequency mode and it does not appear in the normal form Hamiltonian.

\section{Conclusions}\label{sec:Conclusions}

To summarize, in this article we have revisited the discussion of the normal form of a quadratic quantum Hamiltonian describing the linear interaction of quantum harmonic oscillators. We have provided step-by-step instructions to construct a real canonical transformation that can transform a generic quadratic Hamiltonian into its normal form. These tools can be used to unveil the quantum dynamical regimes of a many-mode coupled system as well as to identify the normal modes. This has been illustrated with a minimal example of two harmonic quantum oscillators coupled with a quadratic term in the position operators. Remarkably, this examples already shows the appearance of all the possible normal forms of a quadratic Hamiltonian that correspond to different types of dynamical instabilities.

The discussion of this article is relevant for conservative systems  whose dynamics can be described by a Hamiltonian. In nature, these systems are difficult to find since it is easy to interact with the environment. That is, perhaps, the reason that in the literature one typically finds the discussion of the normal form of quadratic Hamiltonians in the context of celestial mechanics~\citep{Siegel,Meyer}. However, in current quantum optics scenarios, systems in nature can be so well isolated from the environment, that they can also be described by pure Hamiltonians during some relevant time scales. In particular, in the field of quantum nanomechanics~\citep{Aspelmeyer2014}, several mechanical degrees of freedom in mesoscopic systems are so well isolated from the environment that they can be brought and controlled in the quantum regime. We thus believe that it is timely to recall the tools presented in this article to be able to describe the quantum dynamics of current mechanical systems in the quantum regime. These tools can be used to understand and exploit unstable quantum dynamics for a variety of potential applications: from generating squeezing and entanglement via optimal coherent dynamics, to using dynamical instabilities for metrological purposes, something that we plan to study in the future.

We thank J. I. Cirac and B. Kraus for useful discussions. This work is supported by the European Research Council (ERC-2013-StG 335489 QSuperMag) and the Austrian Federal Ministry of Science, Research, and Economy (BMWFW).

\appendix
\section{Generalized Symplectic Orthonormalization}\label{App:gsON}

The generalized symplectic orthonormalization of the gGEVs is an important step in the procedure presented in \secref{sec:NF_Transf} to build the symplectic transformation which transforms the initial Hamiltonian in its normal form.  Here, we describe in detail how such an orthonormalization procedure is carried out. To this end, we start by introducing necessary mathematical tools.

For a given eigenvalue $\lambda$ of the equation-of-motion matrix $\Kquad$, we introduce a matrix of the form~\cite{Meyer,Laub1974}
\begin{equation}\label{Phi}
\Phi_{\lambda} = \sum_{k=1}^{D}\PhiA_k (\Kquad-\lambda\mathbb{1})^{k-1},
\end{equation}
where $\PhiA_k \in \mathbb{C}$ and $D \in \mathbb{N}$. $\Phi_{\lambda}$ is a matrix, but for a given eigenvalue it is completely specified by the vector $\boldsymbol{\Phi}= (\PhiA_1,\dots, \PhiA_D)^T$. Note that acting with ${\Phi_{\lambda}}$ on a gGEV $\xvec \in \egspace(\lambda)$ of rank $D$ gives another gGEV $\Phi_{\lambda}\xvec \in \egspace(\lambda)$ of the same rank~\cite{Meyer,Laub1974}. Additionally, from \eqnref{Phi} we define~\cite{Meyer,Laub1974}
\begin{equation}\label{PhiStar}
\begin{aligned}
\Phi^\star_{\lambda} &= \sum_{k=1}^{D} \tilde\PhiA_k (-1)^{k-1} (\Kquad-\lambda \mathbb{1})^{k-1},
\end{aligned}
\end{equation}
where $\tilde \phi_k = \bar \phi_k$ if $\lambda$ is purely imaginary and $\tilde \phi_k = \phi_k$ otherwise. It is convenient to remark that the matrix multiplication of  two matrices of the form \eqnref{Phi} with the same $\lambda$, say $\Phi_{\lambda} \Theta_{\lambda}$, can be efficiently computed as follows: if $\boldsymbol{\Phi} = (\PhiA_1,\dots,\PhiA_D)^T$ and $\boldsymbol{\Theta} = (\ThA_1,\dots,\ThA_D)^T$, then $\boldsymbol{\Phi}\boldsymbol{\Theta}= (\PhThA_1,\dots,\PhThA_D)^T$, with
\begin{equation}\label{AlgebraProduct}
\quad \PhThA_k=\sum_{l=1}^{k} \PhiA_{l} \ThA_{k+1-l}.
\end{equation}
For an eigenvalue $\lambda$ and two gGEVs $\xvec \in \egspace(\lambda)$ and $\tilde \yvec \in \egspace(-\lambda)$ it is convenient to define~\cite{Meyer,Laub1974}
\begin{equation}\label{Omega}
\begin{aligned}
&\Omega_{\lambda}(\xvec,\tilde\yvec) \equiv \sum_{k=1}^{D} \OmegaA_k (\Kquad - \lambda \mathbb{1})^{k-1},
\end{aligned}
\end{equation}
where  $D$ is the rank of $\xvec$ and
\be 
\OmegaA_k \equiv \spare{(\Kquad - \lambda \mathbb{1})^{D-k}\xvec}^T J\tilde \yvec.
\ee 
Note that for $k=1$, {$\OmegaA_1 \equiv \alpha_{\lambda}(\xvec,\tilde\yvec)$}, see \eqnref{alpha}. $\Omega_{\lambda}(\xvec,\tilde \yvec)$ is a matrix of the form \eqnref{Phi}, and it can be considered as a generalization of the symplectic inner product \eqnref{alpha}~\citep{Laub1974}. As shown in \cite{Meyer,Laub1974}, \eqnref{Omega} has many useful properties, such as $\Omega_\lambda( \xvec, \Phi_{-\lambda} \tilde \yvec) = \Omega_\lambda(\Phi_{\lambda}^\star \xvec, \tilde \yvec)$ and $ \Omega_\lambda(\Phi_\lambda \xvec,\tilde \yvec) = \Phi_\lambda \Omega_\lambda(\xvec, \tilde \yvec)$. Moreover, for imaginary and zero eigenvalue and two gGEVs of the rank $D$, one has that $\Omega(\xvec,\bar\yvec) = (-1)^D \Omega^\star(\yvec,\bar\xvec)$~\cite{Meyer,Laub1974}.

\subsection{Generalized Symplectic Orthonormalization for real and complex eigenvalues, \texorpdfstring{$\mathfrak{c}=1,2$}{c=1,2}}\label{App:gsON-rc}
In this subsection, we describe how to perform a generalized symplectic orthonormalization for $\mathfrak{c}=1,2$. For a given eigenvalue $\lambda_i$, we start with a set of corresponding gGEVs $\lbrace \gvec_{i1}, \dots, \gvec_{im_i}, \tilde \gvec_{i1}, \dots, \tilde \gvec_{im_i} \rbrace $ with ranks $D_{ij}$, where $\gvec_{ij} \in \egspace(\lambda_i)$ and $\tilde \gvec_{ij} \in \egspace(-\lambda_i)$. It is supposed that the gGEVs are ordered such that $\alpha_{\lambda_i}(\gvec_{ij},\tilde\gvec_{ij}) \neq 0$ for $j = 1, \dots, m_i$. In the process of generalized symplectic orthonormalization, this set is transformed to a new set of gGEVs $\lbrace \evec_{i1}, \dots, \evec_{i m_i}, \tilde\evec_{i1}, \dots, \tilde\evec_{i m_i} \rbrace$ which satisfy $\Omega_{\lambda_i}(\evec_{ij},\tilde\evec_{ij'}) = \delta_{jj'} \mathbb{1}$. This set is obtained as follows:
\begin{enumerate}
	\item Normalize $\tilde\gvec_{i1} \rightarrow \tilde\gvec_{i1}/\alpha_{\lambda_i}(\gvec_{i1},\tilde\gvec_{i1})$. Find $\Phi_{\lambda_i}$ of the form of \eqnref{Phi} such that 
	\be 
	\Phi_{\lambda_i}^2 = \Omega_{\lambda_i}(\gvec_{i1},\tilde\gvec_{i1}).
	\ee 
	 The coeficients $\PhiA _k$ defining $\Phi_{\lambda_i}$ can be found by solving a recursive system of equations given by \eqnref{AlgebraProduct}. Then, define
	\be 
	\begin{split}
	\evec_{i1} &= \Phi_{\lambda_i}^{-1} \gvec_{i1}, \\
	\tilde{\evec}_{i1} &= ({\Phi}_{-\lambda_i}^\star)^{-1} \tilde{\gvec}_{i1}.
	\end{split}
	\ee 
	Note that by construction $\Omega_{\lambda_i}(\evec_{i1},\tilde\evec_{i1}) = \mathbb{1}$ and $\alpha_{\lambda_i}(\evec_{i1},\tilde\evec_{i1}) = 1$. 
	\item Redefine the rest of the GEVs as
	\be 
	\begin{split}
	\gvec_{ij} &\rightarrow\gvec_{ij} - \Omega_{\lambda_i} (\gvec_{ij},\tilde{\evec}_{i1})\evec_{i1}, \\
	\tilde{\gvec}_{ij} &\rightarrow\tilde{\gvec}_{ij} - {\Omega}_{-\lambda_i}^\star(\evec_{i1},\tilde{\gvec}_{ij}) \tilde{\evec}_{i1},
	\end{split}
	\ee 
	for $j=2,\dots,m_i$. Note that by construction $\Omega_{\lambda_i}(\evec_{i1},\tilde\gvec_{ij})=\alpha_{\lambda_i}(\gvec_{ij},\tilde\evec_{i1})\id=\mathbb{0}$.
	\item Repeat the above steps for the redefined set $\lbrace \gvec_{i2}, \dots, \gvec_{i m_i},\tilde\gvec_{i2}, \dots, \tilde\gvec_{i m_i} \rbrace$ and keep repeating for $ j=2, \dots,m_i$ until $\Omega_{\lambda_i}(\evec_{ij},\tilde\evec_{ij'}) = \delta_{jj'} \mathbb{1}$ $\forall j,j'$.
\end{enumerate}

\subsection{Generalized Symplectic Orthonormalization for zero eigenvalues, \texorpdfstring{$\mathfrak{c}=3,4$}{c=3,4}}\label{App:gsON-c34}

Let us now describe a process of generalized symplectic orthonormalization for $\mathfrak{c}=3,4$. We start with a set of all gGEVs corresponding to zero eigenvalue, $\lbrace \gvec_{01}, \dots, \gvec_{0 m_0} \rbrace \in \egspace(0)$. In the process of generalized symplectic orthonormalization, this set is transformed to a new set of gGEVs $\lbrace \evec_{01}, \dots, \evec_{0 m_0} \rbrace$, with the property that
\begin{equation}\label{ON-property-c34}
\begin{split}
\Omega_{0}(\evec_{0j},\evec_{0j'}) =& \delta_{jj'} \sigma_{0j} \mathbb{1},\\
\Omega_{0}(\evec_{0j},\evec_{0k})=& \mathbb{0},
\end{split}
\end{equation}
where $j,j'=1,\dots,l_0$, $k = \lzz+1,\dots, m_0$, and $\sigma_{0j}=\pm 1$. The first $\lzz$ gGEVs belong to case $\mathfrak{c}=3$ and have an even rank $D_{0j}$. The $2\nzz=m_0-\lzz$ gGEVs belong to case $\mathfrak{c}=4$ and have an odd rank $D_{0j}$. Note that the number of $\mathfrak{c}=4$ gGEVs is always even. Obtaining the set of gGEVs with the property \eqnref{ON-property-c34} is done in the following way~\cite{Meyer,Laub1974}: 
\begin{enumerate}
	\item Search for a gGEV $\gvec_{0j}$ in the list such that 
	\be\label{eq:Alpha0}
		\alpha_{0}(\gvec_{0j},\gvec_{0j})\neq 0,
	\ee
	and place it at the beginning of the list of gGEVs by $\gvec_{0j}\rightarrow \gvec_{01}$.
	If none of the gGEVs in the list satisfies equation \eqnref{eq:Alpha0} proceed to step four. 	
	Otherwise, define
    \begin{equation} 
    \sigma_{01} = \mathrm{sgn}\spare{\alpha_{0}(\gvec_{01},\gvec_{01})}=\pm 1.\\
    \end{equation}
    Find a matrix $\Phi_{0}$ of the form \eqnref{Phi} such that 
	\be
	\Phi_{0}^2 =  \sigma_{01} \Omega_{0}(\gvec_{01},\gvec_{01}).
	\ee
	$\Phi_{0}$ can be represented by a vector and found using \eqnref{AlgebraProduct}. Define 
	\be 
	\evec_{01} = \Phi_{0}^{-1} \gvec_{01}.
	\ee
	Note that by construction $\Omega_{0}(\evec_{01},\evec_{01})=\sigma_{01} \mathbb{1}$.
	\item Redefine all the remaining gGEVs in the total set $\lbrace \gvec_{02}, \dots, \gvec_{0 m_0} \rbrace$ as 
	\be \label{redefine-c34}
	\gvec_{0j} \rightarrow \gvec_{0j} - \sigma_{01} \Omega_{0}^\star(\evec_{01},\gvec_{0j}) \evec_{01},
	\ee
	for $j=2,\dots, m_0$. By construction $\Omega_{0}(\evec_{01},\gvec_{0j}) =  \mathbb{0}$.
	\item Update the list of gGEV to $\lbrace \gvec_{02}, \dots, \gvec_{0 m_0} \rbrace$ and go to step one.
	\item After the iteration of steps $1-3$ is completed, the list of gGEVs can be ordered as $\{\evec_{01},\ldots,\evec_{0l'_0},\gvec_{0l'_0+1},\ldots \gvec_{0m_0}\}$, where $l'_0\in [0,l_0]$, such that \eqnref{ON-property-c34} is satisfied for $j,j'=1,\dots,\lzz'$ and $k=\lzz'+1,\dots,m_0$. If $l'_0=l_0$, rename $\gvec_{0k} \rightarrow \evec_{0k}$, for $k=l_0+1, \dots, m_0$ and apply to these gGEVs the procedure outlined in \appref{App:gsON-c4}. If $l'_0<l_0$, then there are $l_0-l'_0$ redefined gGEVs with even rank which do not satisfy \eqnref{eq:Alpha0}. In this case, apply only to these vectors the procedure outlined in \appref{App:gsON-c3-extra}.
\end{enumerate}

\subsubsection{There are gGEVs of case \texorpdfstring{$\mathfrak{c}=3$}{c=3} with \texorpdfstring{$\alpha_{0}(\gvec_{0j},\gvec_{0j}) = 0$}{alpha=0}}\label{App:gsON-c3-extra}
Whenever $l'_0<l_0$, the gGEVs in the redefined list $\lbrace \gvec_{0\lzz'+1}, \dots, \gvec_{0 \lzz} \rbrace$ {belong to \texorpdfstring{$\mathfrak{c}=3$}{c=3}, but they} do not satisfy \eqnref{eq:Alpha0}. In this case, for any $j\in [\lzz'+1,\lzz]$, there exist another $j'\in [\lzz'+1,\lzz]$ such that $D_{0j}=D_{0j'}$ and  $\alpha_{0}(\gvec_{0j},\gvec_{0j'}) \neq 0$. This is guaranteed by the nondegeneracy of \eqnref{Omega}~\cite{Meyer,Laub1974}. Consequently, $\lzz-\lzz'$ is even. In this situation, one should proceed as follows:
\begin{enumerate}
	\item Suppose, by reordering the gGEVs if necessary, that $D_{0j}=D_{0j+1}$ and $\alpha_{0}(\gvec_{0j},\gvec_{0j+1}) \neq 0$, for $j=\lzz'+1,\lzz'+3,\dots,\lzz-1$.
	\item Redefine the gGEVs in the following way:
	\be
	\begin{aligned}\label{superposition-c3}
	    \gvec_{0j}\rightarrow \gvec_{0j}+\gvec_{0j+1},\\
	    \gvec_{0j+1} \rightarrow \gvec_{0j}-\gvec_{0j+1}.
	\end{aligned}
	\ee
\end{enumerate}
Since $\alpha_{0}(\xvec,\yvec)$ is linear in both arguments and $\alpha_{0}(\xvec,\yvec)=(-1)^D  \alpha_{0}(\yvec,\xvec)$~\cite{Meyer,Laub1974}, it follows that the superpositions \eqnref{superposition-c3} have a nonvanishing $\alpha_{0}(\gvec_{0j},\gvec_{0j})$:
\be
\begin{aligned}\label{new-alphas-c3}
    \alpha_{0}(\gvec_{0j},\gvec_{0j}) &\rightarrow 2\alpha_{0}(\gvec_{0j},\gvec_{0j+1}),\\
	\alpha_{0}(\gvec_{0j+1},\gvec_{0j+1}) &\rightarrow -2\alpha_{0}(\gvec_{0j},\gvec_{0j+1}).\\
\end{aligned}
\ee
With the redefined gGEVs $\gvec_{0j}$, $j=\lzz'+1,\dots, \lzz$, one can continue with the orthonormalization steps to obtain $\lbrace \evec_{01}, \dots, \evec_{0 m_0} \rbrace$ satisfying \eqnref{ON-property-c34}. From here, one can now proceed to construct the canonical normal form transformation for $\mathfrak{c}=3$, see \secref{sec:NF_Transf_zero-even}.

\subsubsection{Further transformation of the gGEVs \texorpdfstring{$\evec_{0j}$}{} of case \texorpdfstring{$\mathfrak{c}=4$}{c=4}}\label{App:gsON-c4}
Let us analyze the remaining $2\nzz = m_0 - \lzz$ gGEVs in $\mathfrak{c}=4$. To ease the notation, we relabel them as $\lbrace \evec_{01}, \dots, \evec_{0 2\nzz} \rbrace$. In this case, $\alpha_{0}(\evec_{0j},\evec_{0j})=0$  for $ j=1,\dots, 2\nzz$ and it cannot be transformed to some other value like in the case of gGEVs in $\mathfrak{c}=3$. Here we describe a transformation of $\lbrace \evec_{01}, \dots, \evec_{0 2\nzz} \rbrace$ to a new set $\lbrace \fvec_{01}, \dots, \fvec_{0 \nzz}, \hvec_{01}, \dots, \hvec_{0 \nzz} \rbrace$ which satisfies
\be \label{ON-property-c4}
\begin{split}
\Omega_{0}(\fvec_{0j},\fvec_{0j'}) &= \Omega_{0}(\hvec_{0j},\hvec_{0j'}) = 0   \\
\Omega_{0}(\fvec_{0j},\hvec_{0j'}) &= \delta_{jj'} \mathbb{1},
\end{split}
\ee 
for $ j,j'=1,\dots, \nzz$. Obtaining the gGEVs of the form \eqnref{ON-property-c4} is done as follows~\cite{Meyer,Laub1974}:
\begin{enumerate}
	\item Suppose, by reordering the gGEVs if necessary, that $D_{01}=D_{02}$ and $\alpha_{0}(\evec_{01},\evec_{02}) \neq 0$. Such a pair can always be found~\cite{Meyer,Laub1974}. Normalize $\evec_{02} \rightarrow \evec_{02}/\alpha_{0}(\evec_{01},\evec_{02})$. Find a matrix $\Phi_0$ of the form \eqnref{Phi} such that 
	\be 
	\Phi_0^2=\Omega_{0}(\evec_{01},\evec_{02}).
	\ee
	 Make the following transformation: 
	 \be 
	 \begin{split}
	     \evec_{01} &\rightarrow \Phi_0^{-1} \evec_{01}, \\
	     \evec_{02} &\rightarrow (\Phi_0^\star)^{-1} \evec_{02}.
	 \end{split}
	 \ee 
	 By construction, $\Omega_{0}(\evec_{01},\evec_{02})=\mathbb{1}$.
	\item Calculate $\Omega_0(\evec_{01},\evec_{01})$ and $\Omega_{0}(\evec_{02},\evec_{02})$ for the redefined gGEVs.  Find a matrix $\Psi_0$ of the form \eqnref{Phi} that satisfies 
	\be 
	\qquad\Omega_0(\evec_{01},\evec_{01})-2\Psi_0 -\Psi_0^2 \Omega_{0}(\evec_{02},\evec_{02}) = 0.
	\ee
	This equation can be written in the form \eqnref{AlgebraProduct} and solved recursively for the coefficients of the matrix $\Psi_0$.
	\item Define 
	\be 
	\fvec_{01} = \evec_{01} + \Psi_0 \evec_{02}.
	\ee 
	By construction, $\Omega_{0}(\fvec_{01},\fvec_{01})=\mathbb{0}$.
    Repeat step 1 for the pair $\lbrace\fvec_{01},\evec_{02}\rbrace$ to obtain  $\Omega_{0}(\fvec_{01},\evec_{02})=\mathbb{1}$. This step keeps $\Omega_{0}(\fvec_{01},\fvec_{01})=\mathbb{0}$.
	\item Define
	\be
	\hvec_{01} = \evec_{02} - \frac{1}{2} \Omega_{0}(\evec_{02},\evec_{02}) \fvec_{01}.
	\ee
	By construction,  $\Omega_{0}(\hvec_{01},\hvec_{01}) = \mathbb{0}$ and $\Omega_{0}(\fvec_{01},\hvec_{01}) = \mathbb{1}$.
	\item Redefine the rest of the gGEVs as
	\be
	\begin{split}
	\evec_{0j}\rightarrow\evec_{0j} + \Omega_{0}^\star(&\hvec_{01},\evec_{0j}) \fvec_{01}\\ &- \Omega_{0}^\star(\fvec_{01},\evec_{0j}) \hvec_{01},
	\end{split}
	\ee
	for  $j=3,\dots,2\nzz$. By construction, $\Omega_{0}(\fvec_{01},\evec_{0j})=\Omega_{0}(\hvec_{01},\evec_{0j})=\mathbb{0}$.
	\item Repeat the above steps for the redefined set $\lbrace \evec_{03}, \dots, \evec_{0 2n_0} \rbrace$ and keep repeating until \eqnref{ON-property-c4} is satisfied.
\end{enumerate}
With so obtained gGEVs, one can now proceed to construct the canonical normal form transformation for $\mathfrak{c}=4$, see \secref{sec:NF_Transf_zero-odd}.

\subsection{Generalized Symplectic Orthonormalization for imaginary eigenvalues, \texorpdfstring{$\mathfrak{c}=5,6$}{c=5,6}}\label{App:gsON-c56}
Let us now describe a process of generalized symplectic orthonormalization for $\mathfrak{c}=5,6$. For a given eigenvalue $\lambda_i$, we start with a set of all corresponding gGEVs $\lbrace \gvec_{i1}, \dots, \gvec_{i m_i} \rbrace \in \egspace(\lambda_i)$. In the process of generalized symplectic orthonormalization, this set is transformed to a new set of gGEVs $\lbrace \evec_{i1}, \dots, \evec_{i m_i} \rbrace$, with the property that
\begin{equation}\label{ON-property-c56}
\Omega_{\lambda_i}(\evec_{ij},\bar\evec_{ij'}) = \delta_{jj'}\sigma_{ij} \mathbb{1}, \quad  j,j'=1,\dots, m_i,
\end{equation}
where $\sigma_{ij}$ can have the following values:
\begin{equation} \label{sigma}
\sigma_{ij} =
\begin{cases}
\pm 1 & \mathfrak{c}=5,\\
\pm \im &  \mathfrak{c}=6.
\end{cases}
\end{equation}
Obtaining the set of gGEVs with the property \eqnref{ON-property-c56} is done in the following way~\cite{Meyer,Laub1974}: 
\begin{enumerate}
	\item Suppose, by reordering the gGEVs if necessary, that the first gGEV in the list satisfies $\alpha_{\lambda_i}(\gvec_{i1},\bar\gvec_{i1}) \neq 0$. If all are zero, one should skip these steps and proceed to \appref{App:gsON-c56-extra}. Otherwise, define
    \begin{equation} 
    \sigma_{i1} =
    \begin{cases}
    \mathrm{sgn}\spare{\alpha_{\lambda_i}(\gvec_{i1},\bar\gvec_{i1})}= \pm1 &
    \mathfrak{c}=5,\\
    \im \mathrm{sgn}\spare{\alpha_{\lambda_i}(\gvec_{i1},\bar\gvec_{i1})/\im} = \pm \im &
    \mathfrak{c}=6.\\
    \end{cases}
    \end{equation}
	Find a matrix $\Phi_{\lambda_i}$ of the form \eqnref{Phi} such that 
	\be
	\Phi_{\lambda_i}^2 =  \sigma_{i1} \Omega_{\lambda_i}(\gvec_{i1},\bar\gvec_{i1}).
	\ee
	$\Phi_{\lambda_i}$ can be represented by a vector and found using \eqnref{AlgebraProduct}. Define 
	\be 
	\evec_{i1} = \Phi_{\lambda_i}^{-1} \gvec_{i1}.
	\ee
	Note that by construction $\Omega_{\lambda_i}(\evec_{i1},\bar\evec_{i1})=\sigma_{i1} \mathbb{1}$.
	\item Redefine all the remaining gGEVs in the total set $\lbrace \gvec_{i2}, \dots, \gvec_{i m_i} \rbrace$ by 
	\be \label{redefine-c56}
	\gvec_{ij} \rightarrow \gvec_{ij} - \sigma_{i1} \Omega_{\lambda_i}^\star(\evec_{i1},\bar\gvec_{ij}) \evec_{i1},
	\ee
	for $j=2,\dots, m_i$. By construction $\Omega_{\lambda_i}(\evec_{i1},\bar\gvec_{ij}) =  \mathbb{0}$.
	\item Repeat the above steps for the redefined set $\lbrace \gvec_{i2}, \dots, \gvec_{i m_i} \rbrace$ and keep repeating for $ j=2, \dots,m_i' \leq m_i$.
\end{enumerate}
If  $m_i' = m_i$, this completes the procedure. The new set $\lbrace \gvec_{i1}, \dots, \gvec_{i m_i} \rbrace$ satisfies \eqnref{ON-property-c56}. The gGEVs can now be separated into two lists, one for each of the cases $\mathfrak{c}=5,6$. One can now proceed to construct the canonical normal form transformation for $\mathfrak{c}=5,6$, see \secref{sec:NF_Transf_imag-even} and \secref{sec:NF_Transf_imag-odd}.

The case $m_i' < m_i$ happens if $\alpha_{\lambda_i}(\gvec_{ij},\bar\gvec_{ij}) = 0$ for $j>m_i'$, after the redefinition \eqnref{redefine-c56} has been carried out $m_i'$ times. In that case, proceed to \appref{App:gsON-c56-extra}.

\subsubsection{There are gGEVs \texorpdfstring{$\gvec_{ij}$}{gGEVs} with \texorpdfstring{$\alpha_{\lambda_i}(\gvec_{ij},\bar\gvec_{ij}) = 0$}{alpha=0}}\label{App:gsON-c56-extra}

The redefined list of gGEVs $\lbrace \gvec_{im_i'+1}, \dots, \gvec_{im_i} \rbrace$  contains vectors which do not satisfy \eqnref{ON-property-c56}.
In this case,  for any $j\in[m_i'+1, m_i]$, there exist another $j'\in[m_i'+1, m_i]$ such that $D_{ij}=D_{ij'}$ and  $\alpha_{\lambda_i}(\gvec_{ij},\bar\gvec_{ij'}) \neq 0$. This is guaranteed by the nondegeneracy of \eqnref{Omega}~\cite{Meyer,Laub1974}. Consequently, $m_i-m_i'$ is even. In this situation, one should proceed as follows:
\begin{enumerate}
	\item Suppose, by reordering the gGEVs if necessary, that $D_{ij}=D_{ij+1}$ and $\alpha_{\lambda_i}(\gvec_{ij},\bar\gvec_{ij+1}) \neq 0$, for $j=m_i'+1,m_i'+3,\dots,m_i-1$.
	\item Redefine the gGEVs in the following way:
	\be
	\begin{aligned}\label{superposition-c56}
	    \gvec_{ij} \rightarrow \gvec_{ij}+\gvec_{ij+1},\\
	    \gvec_{ij+1} \rightarrow \gvec_{ij}-\gvec_{ij+1}.
	\end{aligned}
	\ee
\end{enumerate}
Since $\alpha_{\lambda_i}(\xvec,\bar\yvec)$ is linear in both arguments and it has the property that $\alpha_{\lambda_i}(\xvec,\bar\yvec)=(-1)^D \bar \alpha_{\lambda_i}(\yvec,\bar\xvec)$~\cite{Meyer,Laub1974}, it follows that the superpositions \eqnref{superposition-c56} have a nonvanishing $\alpha_{\lambda_i}(\gvec_{ij},\bar\gvec_{ij})$:
\be
\begin{aligned}\label{new-alphas-c56}
    \alpha_{\lambda_i}(\gvec_{ij},\bar\gvec_{ij})  \rightarrow
    \begin{cases}
    2\Re\spare{\alpha_{\lambda_i}(\gvec_{ij},\bar\gvec_{ij+1})}, \quad &\mathfrak{c}=5,\\
    2 \im \Im\spare{\alpha_{\lambda_i}(\gvec_{ij},\bar\gvec_{ij+1})}, \quad &\mathfrak{c}=6,\\
    \end{cases}\\
	\alpha_{\lambda_i}(\gvec_{ij+1},\bar\gvec_{ij+1})  \rightarrow
	\begin{cases}
    -2\Re\spare{\alpha_{\lambda_i}(\gvec_{ij},\bar\gvec_{ij+1})}, \quad &\mathfrak{c}=5,\\
    -2 \im \Im\spare{\alpha_{\lambda_i}(\gvec_{ij},\bar\gvec_{ij+1})}, \quad &\mathfrak{c}=6.\\
    \end{cases}
\end{aligned}
\ee
With the redefined gGEVs $\gvec_{ij}$, $j=m_i'+1,\dots, m_i$, one can continue with the orthonormalization steps to obtain $\lbrace \evec_{i1}, \dots, \evec_{i m_i} \rbrace$ satisfying \eqnref{ON-property-c56}.
%
\section{The Bogoliubov Transformation} \label{App:Bog}

In this appendix, we provide the instructions to perform a Bogoliubov real canonical transformation~\cite{Bogoljubov1958,Valatin1958,Maldonado1993} that can be performed when the equation-of-motion matrix $\Kquad$ is diagonalizable and only has purely imaginary eigenvalues. That is, $\Kquad$ has $N_\mathcal{I}$ imaginary pairs $\mathcal{I}_i=\lbrace \lambda_i, \bar \lambda_i\rbrace$, for $i=1,\dots, N_{\mathcal{I}}$ with  algebraic and  geometric multiplicities $a_i=m_i$. One hence needs to use  the generalized symplectic orthonormalization (\secref{App:gsON-c56}) and the instructions for $\mathfrak{c}=6$ (\secref{sec:NF_Transf_imag-odd}) with $D_{ij}=1$. Note that there are $a_i$ eigenvectors corresponding to each $\lambda_i$ and it is not needed to introduce the GEVs.

In the diagonalizable case, the form \eqnref{Omega} introduced in \appref{App:gsON} simply reduces to $\Omega_{\lambda_i}(\xvec,\bar\yvec)= \alpha_{\lambda_i}(\xvec,\bar\yvec) \mathbb{1}$. Moreover, $\alpha_{\lambda_i}(\xvec,\bar\yvec)$ as defined in \eqnref{alpha} further reduces to $\alpha_{\lambda_i}(\xvec,\bar\yvec) = \xvec^T J \bar \yvec$, without dependence on the eigenvalue $\lambda_i$. We will therefore denote it simply as $\alpha (\xvec, \bar \yvec)$. This is the only mathematical object needed in the construction of the normal form transformation in the diagonalizable case. One can show that $\alpha (\xvec, \bar \yvec)=-\bar\alpha (\yvec, \bar \xvec)$ and hence $\alpha (\xvec, \bar \xvec)$ is purely imaginary.
 
Assuming that for each pair $\mathcal{I}_i$ the corresponding eigenvectors $\lbrace \gvec_{i1}, \dots, \gvec_{ia_i} \rbrace\in \egspace(\lambda_i) $ are found, one needs to proceed as follows. For each $\mathcal{I}_i$, do the symplectic orthonormalization as follows:
\begin{enumerate}
    \item Assume that $\alpha(\gvec_{ij},\bar\gvec_{ij})\neq 0$ for $j=1,\dots,a_i$. If this is not the case and $\exists j$ such that $\alpha(\gvec_{ij},\bar\gvec_{ij})=0$, then the number of such eigenvectors is even and there can always be found another $j'$ such that $\alpha(\gvec_{ij'},\bar\gvec_{ij'})=0$, but $\alpha(\gvec_{ij},\bar\gvec_{ij'})\neq 0$~\cite{Meyer,Laub1974}. Then, redefine the corresponding eigenvectors as
    \be
	\begin{aligned}\label{}
	    \gvec_{ij} \rightarrow \gvec_{ij}+\gvec_{ij'},\\
	    \gvec_{ij'} \rightarrow \gvec_{ik}-\gvec_{ij'},
	\end{aligned}
	\ee
	such that $\alpha(\gvec_{ij},\bar\gvec_{ij}),\alpha(\gvec_{ij'},\bar\gvec_{ij'})\neq 0$. See \eqnref{superposition-c56} and \eqnref{new-alphas-c56} for more details.
	\item Define $\asign_{i1} = \im \mathrm{sgn} \spare{\alpha(\gvec_{i1},\bar\gvec_{i1})/\im}=\pm \im$. Note that $-\asign_{i1}\alpha(\gvec_{i1},\bar\gvec_{i1})>0$. Define
	\be
	\evec_{i1} = \gvec_{i1}/\sqrt{-\asign_{i1}\alpha(\gvec_{i1},\bar\gvec_{i1})}.
	\ee
	By construction, $\alpha(\evec_{i1},\bar\evec_{i1})=\sigma_{i1}$.
	\item Redefine the rest of the eigenvectors as
	\be
	\gvec_{ij} \rightarrow \gvec_{ij}-\sigma_{i1}\bar \alpha(\evec_{i1},\bar \gvec_{ij}) \evec_{i1},
	\ee
	for $j=2,\dots,a_i$. By construction $\alpha(\gvec_{ij},\bar \evec_{i1})=0$.
	\item Repeat the above steps for the redefined set $\lbrace \gvec_{i2}, \dots, \gvec_{i m_i} \rbrace$ and keep repeating for $ j=2, \dots,a_i$ until $\alpha(\evec_{ij},\bar\evec_{ij})= 1$ $\forall j$.
\end{enumerate}

The Bogoliubov transformation is $\Tquad=(T_+T_-)$, where the matrices $\Tquad_\pm$ of dimension $2N \times N$ are further split into
\be
T_{\pm} = \pare{T_{1 \pm} \ldots T_{i \pm} \ldots T_{N_\mathcal{I} \pm} },
\ee
where there is a $\Tquad_{i\pm}$ for every imaginary pair $\mathcal{I}_i=\lbrace \lambda_i,\bar \lambda_i\rbrace$ ($i=1,\dots,N_{\mathcal{I}}$) of dimension $2N \times a_i$. The matrices $T_{i \pm}$ are given  by 
\begin{equation} 
\begin{aligned}
&\Tquad_{i+}=\pare{ \uvec_{i1}  \dots  \uvec_{ia_i}},\\
&\Tquad_{i-}=\pare{ \vvec_{i1}  \dots  \vvec_{ia_i}},\\
\end{aligned}
\end{equation}
where
\begin{equation}\label{}
\begin{aligned}
\uvec_{ij} &= \sqrt{2} \Re(\evec_{ij}),\\
\vvec_{ij} &= \im \asign_{ij} \sqrt{2} \Im(\evec_{ij}).
\end{aligned}
\end{equation}
We remark that the matrix $\KNquad=\Tquad^{-1} \Kquad \Tquad$ is not diagonal, but instead it is in the real Jordan normal form (see \tabref{TAB:NormalFormList}). The transformation $\Nquad=\Tquad^T \Mquad \Tquad$  diagonalizes the matrix $\Mquad$, leading to the normal form quadratic Hamiltonian consisting only of independent harmonic oscillators. The Bogoliubov transformation in the bosonic representation is obtained as $\Tbos = G^\dagger \Tquad G$, with $G$ as defined in \eqnref{unitary}.


\end{document}